# Properties of MgB$_2$ bulk

**T A Prikhna**[1,1],

Institute for Superhard Materials of the National Academy of Sciences of Ukraine, 2 Avtozavodskaya Street, Kiev, 04074, Ukraine

E-mail: prikhna@iptelecom.net.ua, prikhna@mail.ru

**Abstract**
The review considers bulk MgB$_2$-based materials in terms of their structure, superconducting and mechanical properties. Superconducting transition temperatures of 34.5–39.4 K, critical current densities of 1.8–1.0·10$^6$ A/cm$^2$ in self field and 10$^3$ in 8 T field at 20 K, 3–1.5·10$^5$ A/cm$^2$ in self field at 35 K, $H_{C2}$ 15 T at 22 K and $H_{irr}$ 13 T at 20 K have been registered for polycrystalline materials. As TEM and SEM study show, dispersed higher borides and rather big amount (5-14%) of oxygen (bonded simultaneously with Mg and B) can be present in the structure even if X-ray pattern contains only reflexes of well crystallized MgB$_2$ with traces of MgO. Materials with such a rather high oxygen content demonstrated high superconducting characteristics and no regularities between a total oxygen content and critical current density have been found. There are no clear regularities between the structural features of the material and its superconducting characteristics ($T_c$, $j_c$, $B_{c2}$, $H_{irr}$), as well as generally accepted mechanisms of the effect of specially introduced additives on these characteristics. At present it is established that nanosized MgB$_{12}$ grains provide effective pinning in polycrystalline material. Besides, additions can introduce the MgB$_2$ structure inducing disorder in lattice sites (for example, C substitution for B). The disorder increases the normal state resistivity, magnetic penetration depth, and the upper critical field, but reduces the transition temperature and anisotropy. It is highly probable that the additives (Ti, Ta, Zr, SiC) together with synthesis or sintering temperature can affect the distribution of oxygen and hydrogen in the material structure as well as the formation of grains of higher borides, thus influencing superconducting properties. The superconductivity of materials with matrix close to MgB$_{12}$ in stoichiometry ($T_c$=37 K) has been defined. The highest mechanical properties achieved for bulk MgB$_2$ –based materials are as follows: the Vickers hardness under a 148.8 N-load H$_v$=10.12±0.2 GPa and fracture toughness under the same load K$_{1C}$=7.6± 2.0 MPa m$^{0.5}$, Young modulus E=273 GPa. The bulk MgB$_2$–based materials can be used for cryogenic machine-building (electromotors, pumps, generators), for magnetic shielding and fault current limitation, in microwave devices.

**Contents**

---

[1] prikhna@iptelecom.net.ua, prikhna@mail.ru

**1. Introduction**

Interest in the MgB$_2$-based materials (known from the early 1950's, whose superconducting (SC) properties were defined in 2001 [1]), despite the comparatively low SC transition temperature (39-40 K [2]), can be explained on the one hand, by the simple hexagonal crystal structure (a = 0.3086 nm, b = 0.3524 nm) and large coherence lengths ($\xi_{ab}(0)$ = 3.7 ÷ 12 nm, $\xi c(0)$ = 1.6 ÷ 3.6 nm) of MgB$_2$, its low density (theoretical density of MgB$_2$ $\rho$ = 2.55 g/cm$^3$) [3], transparency of grain boundaries to current [3, 4], and the possibility to achieve high critical current densities and fields in polycrystalline material, in particular, by simpler and cheaper preparation technique (than in the case of high temperature superconductors (HTSC)), aptitude of the materials for both large scale applications and electronic devices, and on the other hand by the intensive development of technologies that use liquid hydrogen as an alternative fuel for motor, water and aviation transports as well as for the transmission of electrical energy for long distances, because the boiling temperature of liquid hydrogen (20 K) can be working temperature for MgB$_2$-based superconductive materials.

The study of MgB$_2$ energy gap showed its two-gap nature [5-16]. The values of gaps energy allow the conclusion that MgB$_2$ combines features of I and II types superconductors and thus, are belong the so-called 1.5-type superconductors [17]. The fact that both gaps close at the same transition temperature [7-10, 15, 18-20] can be the evidence that the first low-temperature gap is induced by the second (larger) one, and in the moderate or high magnetic fields it behaves as II-type superconductor. For the efficient application of MgB$_2$-based superconductors further improvement of the in-field $J$c is needed. There are two classes of properties that can be responsible for the $J$c and might be termed 'intrinsic' and 'extrinsic' [21]. According to classification given in [21], intrinsic (i.e. intragranular) properties of polycrystalline MgB$_2$ are $H_{c2}$ and flux pinning (or, alternatively, intragrain $J$c). Extrinsic properties (connectivity and porosity) improvement can be attained by substantial across-the-board increases in $J$c and can accompany an increase in the superconductor-material's effective cross-sectional area for the conduction of transport current (the possibility of 'cross-sectional deficiency' was first recognized by Rowell [22, 23]). Any kind of disorder potentially changes the properties of MgB$_2$. Disorder can be introduced in a controlled way by doping or irradiation, but often arises from the preparation conditions [4]. Disorder generally decreases the transition temperature [4]. It has been suggested that intrinsic properties are affected by: (i) macroscopic particles that contribute to lattice distortion enhance both $\pi$ and $\sigma$ scattering, (ii) disorder in the Mg sublattice (e.g. by Al addition) can increase the $\pi$ scattering, (iii) oxygen or carbon when substituting for B is expected to provide strong $\sigma$ scattering [21, 24]. Superconductivity in the $\pi$-band is suppressed at high magnetic fields, where the $\sigma$-band determines the magnetic properties and MgB$_2$ behaves as a single-gap superconductor [4]. The $\pi$-band contributes significantly to the condensation energy and to the superfluid density only at low magnetic fields (below about 1 T in clean materials) [4]. Clean grain boundaries are no obstacles for supercurrents in MgB$_2$ [4, 25-31]. This advantage compared to high temperature superconductors allows simple preparation techniques, but, the connections between the grains remain delicate, since dirty grain boundaries potentially reduce the critical currents [4, 22, 32]. In general, current transport in polycrystalline MgB$_2$ samples tends to be partially blocked by pores and grain-boundary precipitates so that the effective transport cross-sectional area is less than the conductor's geometrical cross-sectional area [21]. As it was summarized in [4] that bulk pinning was weak in MgB$_2$ single crystals and no indications for strong bulk pinning were observed in thin films. There were concluded that grain boundaries seemed to be the dominant pinning centres in thin films and high critical current densities close to 15% of the depairing current density were reached [4]. According to this is the expected maximum for loss-free supercurrents in the case of optimized pinning and the nanosized MgB$_2$ grains provide enough grain boundaries to reach the theoretical limit. Pinning seems to be similarly strong in wires, tapes or bulk samples. The observations show that grain boundary pinning may be dominant in sintered bulk MgB$_2$ [33]. The connectivity reduces the critical currents by a factor of about five in today's best 'high field conductors' [4]. The anisotropy further suppresses the critical currents at high magnetic fields and is successfully described by a percolation model. The introduction of disorder increases the upper critical field and reduces its anisotropy, leading to higher currents at high magnetic fields [4].

The introduction of dopants can have one or more effects, which may combine an increase in the bulk pinning strength [21]: (1) by increasing the crystal's $H_{c2}$ and $H_{irr}$; (2) by forming a wide distribution of point pinning centers each described by an elementary pinning force $f_p$; (3) by producing localized lattice strains, which also contribute to flux pinning. The magnitude and behavior of $J_c$ in an applied field $H$ is described by a bulk pinning function, $F_p$, which represents an appropriate summation of the elementary pinning forces, $f_p$ [21, 34–36].

$$F_p = H_{c2}^m f(h), \text{ where } h = H/H_{c2} \qquad (1)$$

The dopants added to increase $H_{c2}$, although are not pinning centers themselves, serve to increase the measured $F_p$ of already-pinned samples [21].

Despite the comparatively simple lattice structure of MgB$_2$ compound, to find correlations between MgB$_2$-based material structural features and its superconducting properties is a very complicated task. It can be explained by difficulties originated from the necessity to detect the amount and distribution of boron, to analyze the boron-containing compounds [37] and from the complicities connected with analyzing nanostructural materials, often porous, which in addition are disposed towards easy bonding of oxygen and hydrogen.

The present review is aimed at analyzing the reasons and regularities of origination of structural inhomogenities in bulk MgB$_2$-based material, which can affect its superconducting characteristics (critical current density in magnetic field, in particular); discussions of mechanical properties and material stability, because the work in magnetic fields and under the conditions of thermocycling imposes definite requirements on the materials. The first results connected with the application of MgB$_2$-based bulk materials in cryogenic devices are as well under the consideration.

## 2. Composition and analysis of MgB$_2$-based material structure.

*2.1. Typical structure of MgB$_2$-based bulk materials.*

The method of X-ray phase analysis widely applied in material sciences is, unfortunately, not sufficiently informative in the case of MgB$_2$-based materials [37-43]. Figures 1–2 show the results of X-ray, SEM studies (using quantitative EDS analysis) and dependences of $J_c$ of the samples prepared under high (2 GPa) pressure from Mg and B (figures 1, 2b, d, e) and from previously synthesized MgB$_2$ (figures 2a, c, e) [39, 40, 41]. Densities of synthesized and sintered under high pressure samples are very high (they are near theoretically dense). The X-ray pattern of synthesized material (figure 1a) contains reflexes of well crystallized magnesium diboride, small amount of MgO, MgH$_2$ and Mg. In the X-ray pattern in figures 2c, d the MgH$_2$ reflexes are practically absent. It can be due to the absence of H$_3$BO$_3$ or other hydrogen-contained admixtures in the initial magnesium diboride or boron (fresher prepared) or due to high sintering temperature (1000 $^o$C) at which the solubility of hydrogen in material decreases and it evaporates. It was observed by authors of [42, 43] that the presence of MgH$_2$ causes superconducting characteristics of MgB$_2$-based materials to reduce, $J$c, in particular. As the SEM study showed, high-pressure sintered and synthesized samples (figures 1c and 2a, b) contained dispersed inclusions of higher borides (with near MgB$_{12}$ stoichiometry) and that there are no reflexes in X-ray pattern which can be ascribed to this phase. It should be mentioned that in the early papers of Prikhna et. al., for example [42, 43], the composition of "black" Mg-B inclusions was miscounted and later the same results of SEM (EDX-analysis) were recalculated, and near MgB$_{12}$ composition of "black" inclusions for all studied high-pressure manufactured materials was confirmed. The reflex at 2Θ=26.8$^o$ marked "X" in figures 2c, d can be attributed to hexagonal BN and for a long time it was considered to be the reflex of BN because MgB$_2$-based materials under high pressure conditions have been manufactured in contact with hexagonal BN. Despite the fact that BN cannot deeply penetrate into MgB$_2$ sample, a thin layer of BN is usually formed on the surface of MgB$_2$ sample and to remove it is rather difficult (it can be removed by grinding).

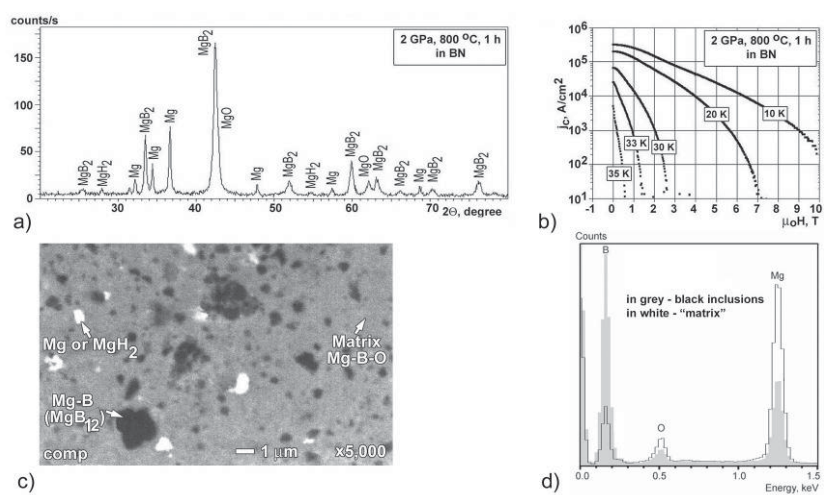

**Figure 1.** Characteristics of the sample synthesized at 2 GPa, 800 $^o$C for 1 h from Mg and B (amorphous, MaTecK 95-97% purity) [39]: (a) X-ray pattern; (b) critical current densities ($j_c$) at different temperatures vs. magnetic field (μ$_0$H) variation estimated by magnetic method; (c) COMPOsitional or backscattering electron image, (d) energy-dispersive spectra (gray-colored spectrum is the spectrum of the "black" inclusions with near MgB$_{12}$ stoichiometry, white-colored spectrum is the spectrum of the "matrix" Mg-B-O phase (Mg and B in approximately MgB$_2$ proportion) of the structure shown in figure 1c). (c) and (d) were obtained by SEM Superprobe JEOL JXA 8800L.

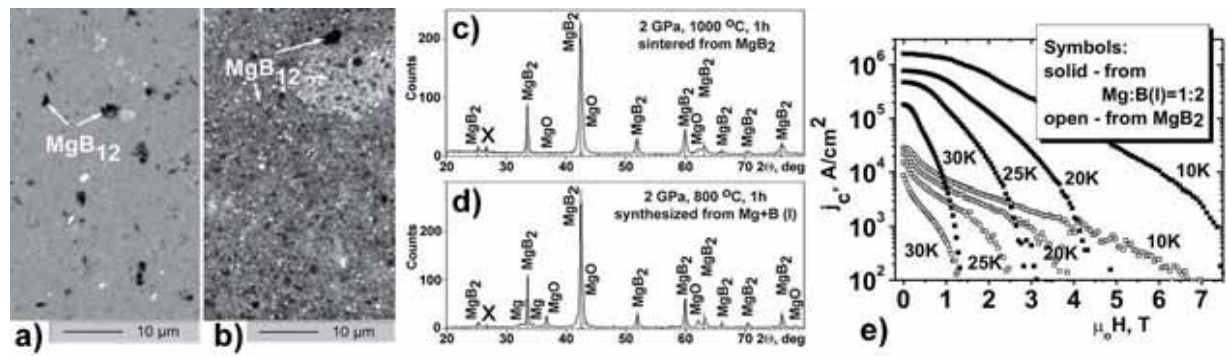

**Figure 2.** (a), (b) structure of the samples obtained by SEM in COMPOsitional contrast using SEM Superprobe JEOL JXA 8800L[40, 41]:
(a) sintered from MgB$_2$ (H.C. Starck, 10 $\mu$m grain size with 0.8% of O) at 2 GPa, 1000 $^o$C, 1 h; black inclusions contain 86 wt% B, 13 wt% Mg, and 1.0 wt% O (near MgB$_{12}$ stoichiometry); grey matrix that contains 48 wt% B, 45 wt% Mg, and 7.0 wt% O ( near MgB$_2$ stoichiometry); very bright small inclusions (seems from material of milling bodies, which were present in the initial powders) containing Zr, Nb and O (ZrO$_2$?);
(b) synthesized from Mg and B (amorphous, H.C. Starck, 1.4 µm grain size with 1.9 % O) taken in the 1:2 ratio at 2 GPa, 800 $^o$C, 1 h; gray matrix contains 40 wt% B, 50 wt% Mg, and 10 wt% O and light matrix contains 25 wt% B, 50 wt% Mg, and 25 wt% O; very bright small inclusions are Mg or MgO or ZrO$_2$ from milling (the latest were in the initial boron);
(c),(d) –X-ray pattern of the samples shown in Figs. 2a,b;
(e) dependences of critical current density, $j_c$, on magnetic fields, $\mu_o$ H, at different temperatures of the samples shown on Figs 2a, b: open symbols –sintered from MgB$_2$ material and solid symbols-synthesized from Mg and B taken in the 1:2 ratio.

Besides, the intensity of this reflex was rather small and it was easily considered to be originated by hexagonal BN admixture. The careful X-ray and SEM studies of materials prepared from mixtures with higher amount of boron (from Mg:B=4 up to Mg:B = 1:20) show that the intensity of the 2$\Theta$=26.8$^o$ reflex usually increases with the increase of amount of higher borides in material structure, besides, SEM EDX analysis shows the absence of nitrogen in the materials [42].

The presence of MgB$_{12}$ in MgB$_2$ –based material structure was mentioned by several researchers [40, 44-46]. In [46] it was even pointed out that if the elements boron and magnesium are used for preparation, a higher Mg-boride, MgB$_{12}$, is always formed. The sizes of B- enriched phase grains in high-pressure (2 GPa) synthesized and sintered MgB$_2$ materials varied from 10 µm to approximately 20 nm. It should be mentioned that in materials prepared under high-pressure (2 GPa) conditions the stoichiometry of B- enriched phase grains distributed in matrix was near MgB$_{12}$, while in material prepared at hot pressing (pressure of 30 MPa) it was near MgB$_6$–MgB$_7$. The presence of B-rich phases and higher borides MgB$_4$ and MgB$_7$ in MgB$_2$ wires, tapes and bulk HIPed (high isostatic pressed) material reported in [37, 47] (see, figure 3). It was defined by the TEM study [37] that B-rich phases with 200–300 nm grains may occupy approximately 1.4% of volume fraction in an undoped MgB$_2$ material prepared from a mechanically alloyed Mg + 2B mixture and 2.7% of volume fraction in the core of rectangular wire prepared from the mixture of Mg + 2B + 11.3 mol% SiC. Giunchi G. et al. [48] established the presence of phase with near MgB$_{13}$ stoichiometry in MgB$_2$ bulk material obtained by liquid Mg infiltration into boron powder. The existence of a binary compound MgB$_{12}$ with theoretical composition 15.5 wt% of B and 84.5 wt % of Mg was claimed in 1955 by Markovski et al. [49]. This phase was mentioned on Mg-B phase diagram at 4.5 GPa and 2 GPa [50, 51]. In 2002 Brutti et al. [52] denied the existence of MgB$_{12}$. Recently V Adasch et al. [53, 54] synthesized MgB$_{12}$ single crystals from the elements in a Mg/Cu melt at 1600 $^o$C and shown that MgB$_{12}$ crystallizes orthorhombic in space group Pnma with $a$ =1.6632(3) nm, $b$ =1.7803(4) nm and $c$=1.0396(2) nm. Schmitt R [44] synthesized polycrystalline MgB$_{12}$ at 1200 °C from Mg und B with trigonal structure $a$= 1.1014(7) nm, $c$ = 2.4170(2) nm. The standards X-ray data for MgB$_{12}$ up to now are absent in the database and the data from literature are very different and contradictive. The MgB$_{12}$ phase may crystallize in the shape (habit) of approximately hexagonal grains (or at least they may have approximately hexagonal cross-section) in MgB$_2$ matrix [55] (figure 4).

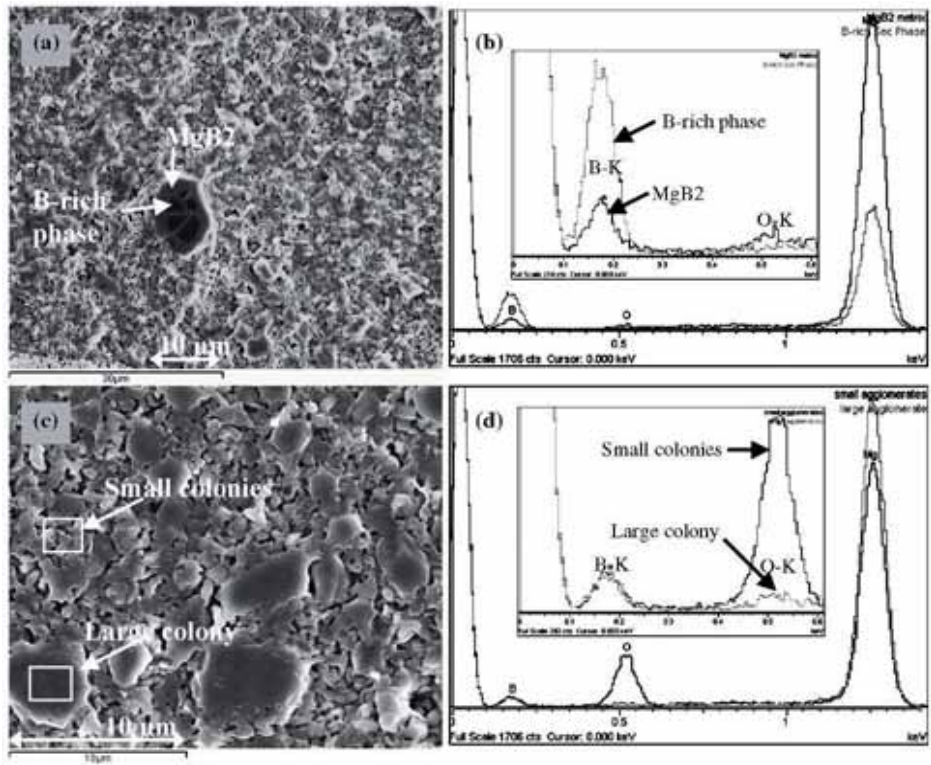

**Figure 3.** SEM images of the cross-section of a tape showing (a) B-rich phase and the MgB$_2$ matrix and (c) large and small MgB$_2$ colonies. The SEM-EDX spectra obtained from the regions indicated in (a) and (c) are shown in (b) and (d), respectively. Portions of the spectra around B K$\alpha$ and O K$\alpha$ peaks are shown magnified in the insets [47].

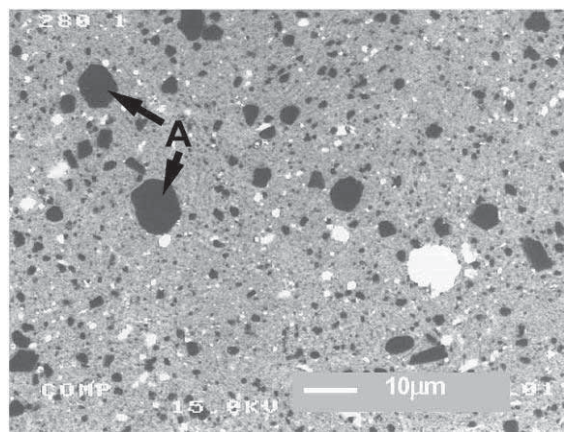

**Figure 4.** Structure of MgB$_2$-based material synthesized from amorphous boron and magnesium chips taken in Mg:B=1:2 ratio in contact with Ta foil at 2 GPa, 800 °C, 1 h (composition image) shows that inclusions of black phase with near MgB$_{12}$ stoichiometry may crystallize in hexagonal habit or have near hexagonal cross-section (see, for example, inclusion marked by "A") [55].

**Table 1.** SEM-EDX quantification of the MgB$_2$ matrix and B-rich phases in large (~5 μm) colonies and small (~1 μm) colonies [47].

|              | B (at.%) | O (at.%) | Mg (at.%) | B (at.%)/Mg (at.%) |
|--------------|----------|----------|-----------|---------------------|
| MgB$_2$      | 66.01    | 0.91     | 33.08     | 2.00                |
| B-rich phase | 89.20    | 0.19     | 10.61     | 8.41                |
| Large colony | 65.21    | 1.14     | 33.64     | 1.94                |
| Small colonies | 56.75  | 17.36    | 25.89     | 2.19                |

The SEM-EDX spectra acquired from the regions indicated in figures 4a and 4c are shown in figures 3b and 3d and the elemental quantification is given in Table 1 [47]. It can be seen that the cores of the large colonies have a B/Mg mole fraction ratio close to two and the oxygen mole fraction is ~1 at.%. In the spectrum obtained from a region containing many small MgB$_2$ colonies, a large (~16 at.%) oxygen mole fraction is found. The dark phases shown in figure 3a were found to be boron-rich phases with boron to magnesium mole fraction ratio of about 8.

The SEM (EDX) study of big amount of high-pressure (2 GPa) synthesized and sintered samples without and with additions shown that the amount of oxygen in the matrix phase of MgB$_2$ is higher than that in MgB$_{12}$ (figure 5) and that it was no regularities found (figure 6) between the amount of oxygen present in the material and its amount in the initial boron or magnesium diboride, as well as with values of critical current density (for example, in 1 T field) [56].

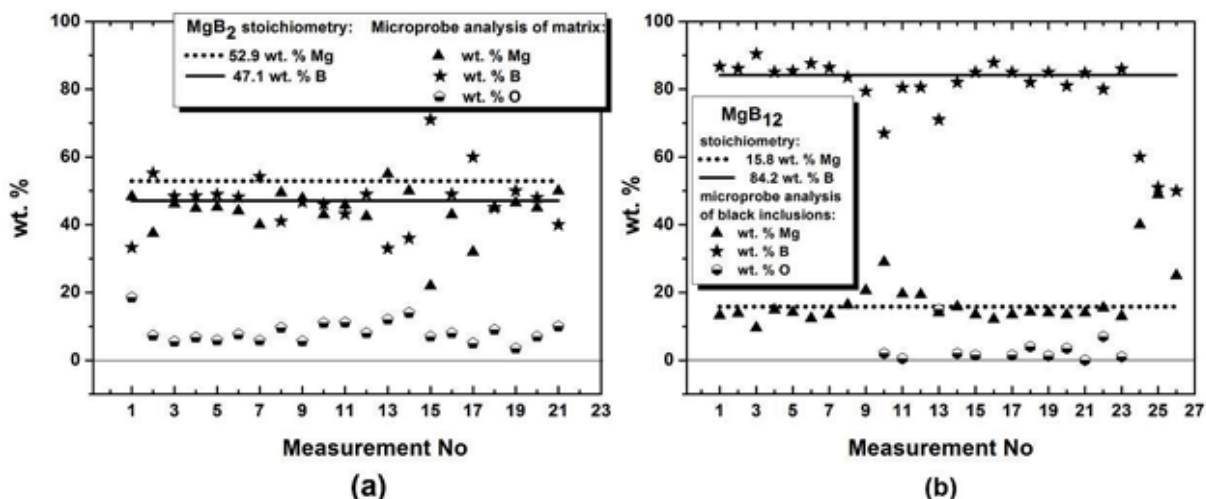

**Figure 5.** Statistical data of SEM microprobe analyses (weight amount of B, Mg and O) [40] of different high-pressure synthesized from Mg and B taken in 1:2 ratio and sintered from powdered MgB$_2$ samples (a) of B-enriched inclusions located in material matrix, compositions of which were very close to MgB$_{12}$ and (b) of matrix phases with near MgB$_2$ stoichiometry. The lines depicting the amounts of magnesium (dot line) and boron (dance thick line) that correspond to the ideal MgB$_{12}$ or MgB$_2$ stoichiometries, respectively, are plotted for comparison.

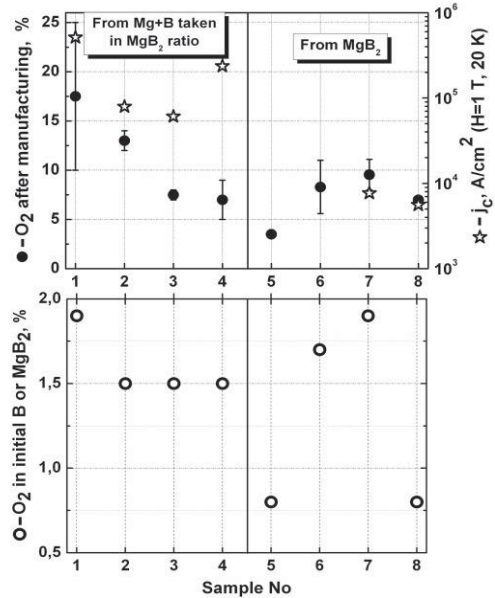

**Figure 6.** The amount of oxygen (wt.%) obtained by SEM microprobe analysis in the initial boron or MgB$_2$ powder (o) and in the prepared (•) materials (synthesized from Mg and B taken in 1:2 ratio or sintered from MgB$_2$) and critical current density (★) in 1 T field at 20 K estimated using VSM in different samples manufactured at 2 GPa for 1h at 800 °C (Nos 1, 2, 5), 900 °C (Nos 3, 6, 7) and 1000 °C (Nos 4, 8). Superconductive phase was present in samples 5 and 6, but the shape of hysteresis loops witnessed the absence of connectivity between the superconductive grains, thus as a whole the materials were not superconductive. The sample 1 prepared from boron with average grain size of 1.4 μm, samples 2, 3, 4 from 4 μm boron and samples 5 and 8 from 10 μm, sample 6 from 9 μm, sample 7 from 1.4 μm MgB$_2$ [56].

A continuous distribution of $T_c$ and $B_{c2}$ (or $H_{c2}$) can be expected for the grains of inhomogeneous polycrystalline materials [57], as indicated by an increased superconducting MgB$_2$. Such inhomogeneities change the temperature dependence of $B_{c2}$ [4, 58]. One of the arguments favouring grain boundary pinning is provided by [59], where a clear correlation between the grain size and $J_c$ was summation of pinning forces found [4]. An increase in the number of grains (and boundaries) by a factor of 6 led to an enhancement of $J_c$ by nearly the same factor (5.5), as expected from a direct summation of pinning forces [4]. From the TEM dark–field images the authors of [47] estimated the size of a typical MgB$_2$ grain of the *ex-situ* tape at being between 0.5 and 1 μm, but many MgB$_2$ grains of size between 30 and 200 nm were also observed (figure 4). In [37] the MgB$_2$ grain size of about 500 nm was mentioned for *ex-situ* commercial, 14-filament MgB$_2$ tape produced over long length scales and grains less than 50 nm for *in-situ* prepared material from a mechanically alloyed Mg and B mixture (with Jc in 6 T field at 4.2 K of $1.4 \cdot 10^4$ A/cm$^2$ for the *ex-situ* and of $6 \cdot 10^4$ A/cm$^2$ for the *in-situ* materials). However, the attempts to find correlations between the average sizes of crystallites, synthesis or sintering temperature and critical current density, $j_c$, for high-pressure manufactured MgB$_2$-based materials failed (Table 2) [41], may be because this variation was negligibly small in the materials prepared by this method under different temperatures and other factors had stronger influence on Jc. The average crystallite sizes in [41] were calculated from line broadening in X-ray diffraction pattern by the standard program in accordance with the following:

$$\text{Crystallite size} = \frac{K \cdot \lambda}{W_{\text{size}} \cdot \cos\theta} \quad \text{with} \quad W_{\text{size}} = W_b - W_s \tag{2}$$

**Table 2.** Critical current density, $j_c$, vs. relative average grain size of crystallites of high-pressure sintered (HPS) from MgB$_2$ and synthesized (HPS) from Mg and B taken in 1:2 ratio materials [41].

| HPS under 2 GPa for 1 h at T$_s$, °C | average crystal size | lattice parameters | | $j_c$, kA/cm$^2$ at 10 K | | $j_c$, kA/cm$^2$ at 20 K | |
|---|---|---|---|---|---|---|---|
| | | $a$ (nm) | $c$ (nm) | 0 T | 1 T | 0 T | 1 T |
| From MgB$_2$ (10 μm and 0.8 % of O) | | | | | | | |
| 700 | 19.7 nm | 0.30805 | 0.35188 | - | - | - | - |
| 800 | 18.8 nm | 0.30822 | 0.35212 | - | - | - | - |
| 900 | 18.5 nm | 0.30820 | 0.35208 | 56 | 14 | 36 | 8 |
| 1000 | 25.0 nm | 0.30797 | 0.35200 | 28 | 8 | 19 | 5 |
| From Mg chips and B (4 μm, 1.5 % O) mixed and milled in 1:2 ratio | | | | | | | |
| 800 | 15.0 nm | 0.30747 | 0.35188 | 245 | 142 | 138 | 79 |
| 900 | 21.0 nm | 0.30819 | 0.35174 | 205 | 136 | 128 | 61 |
| 1000 | 37.0 nm | 0.30808 | 0.35192 | 485 | 364 | 360 | 237 |

where $W_{size}$ is the broadening caused by small crystallites; $W_b$ is the broadened profile width; $W_s$ is the standard profile width (0.08 °); $K$ is the form factor; $\lambda$ is the wavelength.

Crystallographic parameters of MgB$_2$ prepared under 2 GPa pressure in 700-1000 °C temperature diapason varied for the sintered from MgB$_2$: parameter $a$ from 0.30797 to 0.30840 nm and parameter c from 0.35192 to 0.35212 nm and for the synthesized from Mg and B: parameter a from 0.30747 to 0.30819 nm and parameter c from 0.35174 to 0.35192 nm. In the review of Buzea [3] for MgB$_2$ are given parameters: $a = b = 0.308468(8)$ nm, $c = 0.35244(1)$ nm and Jones and March [60] in 1953 characterizing MgB$_2$ wrote that it crystallizes in the hexagonal AlB$_2$ type in the space group P6/mmm with the lattice parameters $a = 0.30834(3)$ nm, and $c = 0.3522(2)$ nm [45]. Despite the rather high oxygen content in high-pressure manufactured material and wide diapason of variation of its amount 5-14 wt%, the lattice parameters and position of X-ray reflexes are practically the same as approved for pure MgB$_2$, what may be the result of their insensibility for oxygen incorporation into MgB$_2$ structure.

Long straight dislocations were seen in MgB$_2$ colony of dense grains in MgB$_2$ tapes [37] with an average spacing of ~0.1 μm, which corresponds to a dislocation density of $1 \times 10^{10}$ cm$^{-2}$.

In the overview of Birajdar et al. [37] the following conclusions concerning methods of MgB$_2$ structure investigation have been made: (i) MgB$_2$, X-ray phase analysis is not able to detect B-rich secondary phases like MgB$_4$ and MgB$_7$ due to poor diffracted signals because of the low x-ray atomic scattering factor of B. Also, the spatial distribution of chemical phases cannot be studied using XRD. (ii) with rapid developments in the technology of energy-dispersive x-ray spectroscopy (EDX) detectors, digital pulse processors and energy filters, electron beam techniques are able to detect light elements with high signal to background ratio. (iii) chemical mapping techniques like EDX elemental mapping and electron spectroscopic imaging (ESI) have become powerful imaging tools in commercial electron microscopes. Therefore electron beam spectroscopy and chemical mapping techniques are suitable for the quantitative chemical analysis of light elements like B, C and O.

It was pointed out in [37] that quantitative EDX analysis of B is difficult because of the higher absorption of low energy x-rays (B K$\alpha$ at 183.3 eV) in the material and the small fluorescence yield of B K$\alpha$ X-rays. The reliable standard EPMA analysis (electron probe micro-analysis) of B established by [61, 62] is time consuming. Therefore, a standard-less, quick and reliable method of B quantification has been proposed in [37] using EDX (energy-dispersive x-ray spectroscopy) in SEM (scanning electron microscope), for the analysis of MgB$_2$ wires and tapes with calibration of the SEM-EDX detector efficiency for B K$\alpha$ X-rays by the manufacturer [63] using metallic boron as standard. It was found by authors of [37] that B/Mg mole fraction ratios increase with increase in sample tilt with respect to (towards) the detector what was explained by absorption of low-energy B K$\alpha$ x-rays in the specimen. And because of this it was concluded [37] that for quantitative analysis the amount of sample tilt with respect to the plane perpendicular to the electron beam has to be checked precisely.

The structure of MgB$_2$-based material with high critical current densities can be multiphase and contain rather high amount of oxygen (5-14 wt %) bonded with Mg and B; boron-enriched (as compared to MgB$_2$ stoichiometry) inclusions or grains of higher borides which are practically free from oxygen, as well as some amount of MgO and free Mg. MgB$_2$-based material may contain MgH$_2$, but usually with the increase of amount of this phase critical current density of material reduces. In dance high-pressure manufactured material amount of MgO did not vary essentially with synthesis or sintering temperature at least in 700-1100 $^o$C temperature range, it was not found correlations between sizes of crystallites and critical current density (may be due to the small sizes of crystallites and narrow diapason of their variation), as in the case with material prepared under ambient pressure (for which an increase in the number of grains (and boundaries) led to an enhancement of $J$c), as well as between total amount oxygen in the high-pressure manufactured material and its SC characteristics.

*2.2. Oxygen distribution in the MgB$_2$ structure*

The admixture of oxygen in the MgB$_2$ material structure has been considered to be harmful for superconducting properties of MgB$_2$ because of the formation of MgO, appearance of which makes boundaries between grains "dirty", which leads to a decrease of effective cross-sectional area through which current can flow or reduce the "connectivity" and thus affect the critical current density decrease. But the latest experience [37, 47, 64-68] has shown that the superconducting properties can be improved by the distribution of oxygen in the MgB$_2$ structure in a certain way. Eom et al. [64] have established that in thin MgB$_2$ films the substitution of oxygen for boron in the boron layers (to form films with a c-axis parameter of 0.3547 nm, which is larger than that for bulk material: 0.3521 nm) leads to a lower T$_c$ but to a steeper slope of d$H_{c2}$/d$T$ both in the parallel and perpendicular magnetic field than that for films with normal parameters. Also, the authors supposed that additional co-pinning by the non-superconducting MgO particles can contribute to the total pinning force.
Using high-resolution transmission electron microscopy (HREM) Liao et al. [67, 68] have shown that the oxygen substitution occurs in the bulk of MgB$_2$ grains to form coherently ordered MgB$_{2-x}$O$_x$ precipitates from about 5 up to 100 nm of size and that such precipitates can act as pinning centers, thus increasing the critical current density. These precipitates are formed due to the ordered replacement of boron atoms by atoms of oxygen and are of the same basic structure as the MgB$_2$ matrix but with composition modulations. No difference in the lattice parameters between the precipitates and the matrix can be detected in conventional electron diffraction patterns. However, extra satellite diffraction spots are seen in some directions implying the structural modulation nature of the precipitates. The precipitates have the same orientation as the MgB$_2$ and the replacement of boron by oxygen makes the precipitates stronger in electron scattering. The periodicity of oxygen atom ordering depends on the concentration of oxygen atoms in the precipitate and first of all occurred in the (010) plane (figure 7) [68].

Figure 8a shows the presence of oxygen-containing precipitates of different sizes in the material. The fact that large precipitates contain mainly magnesium and oxygen, while small precipitates contain all three elements (magnesium, boron, and oxygen), figure 8b, implies that the precipitate composition varies with the precipitate size. The oxygen content increases, while the boron content decreases with increasing precipitate size, indicating the substitution of oxygen for boron. It is interesting that Mg(B,O)$_2$ has the same lattice structure as the MgB$_2$, similar lattice parameters and the crystallographic orientation (figures 8c–d). Estimating the influence of the precipitates on the pinning process, Liao et al. [68] have come to a conclusion that the precipitates of radius larger than ~ 5 nm (or diameter larger than ~ 10 nm) are relevant pinning centers in MgB$_2$. The following mechanism of the precipitate (which includes oxygen) formation has been proposed in [68]: oxygen, dissolved in MgB$_2$ at a high temperature, is later forced out to form Mg(B,O)$_2$ precipitates due to its lower solubility at lower temperatures. A long-term exposure to oxygen at high temperatures results in the transformation of Mg(B,O)$_2$ precipitates to MgO with a little change in precipitate sizes. The density of the distribution of the precipitates of sizes 10–50 nm has been $10^{15}$/cm$^3$, and of size 50–100 nm has been $10^{14}$/cm$^3$ [68]. The unit cell structure of the inclusions does not practically differ from that of the MgB$_2$. So, in parallel with the low X-ray atomic scattering factor of B the aforesaid can be the explanation of the difference in results between X-ray and SEM-EDX examinations of MgB$_2$-based materials.

The oxygen-enriched regions and MgO precipitates of size 15–70 nm were found in MgB$_2$ grains of tapes [47]. The high O/Mg mole fraction ratio (0.23 for precipitates and 1.64 for an O-rich region) gave to the authors of [47] grounds to consider the precipitate as being MgO and the O-rich region being a mixture of MgO and Mg(OH)$_2$.

As it is shown in [41, 66] for high-pressure synthesized material, the increase of synthesis temperature (at least from 800 $^o$C to 1050 $^o$C) induces the segregation of oxygen (figures 9a, b) with the formation of Mg-B-O inclusions oxygen-enriched as compared with the material matrix and thus leads to the reduction of the amount of oxygen in material matrix. So, there occur in some sense the process of grain boundary cleaning and an increase of connectivity. Such structural variations result in the increase of critical current density in low magnetic fields, but in high magnetic fields $J_c$ somewhat decreases (figures 10a, c). It should be mentioned that in parallel with this process a decrease in the amount of B-enriched phases is observed with increasing synthesis temperature (the B-enriched phases can affect critical current density), which is discussed in details here later (section 2.3.).

Additions of Ti or Ta (figures 9g, h) and obviously SiC (figures 11a, b) in combination with an increase of the synthesis temperature make the process of oxygen segregation (figure 9n) more pronounced. In the gray matrix of an MgB$_2$-based material with 10 wt % of Ti addition synthesized at 800 $^o$C (figure 9c) the amount of oxygen was about 8% and in that synthesized at 1050 $^o$C (figures 9h, k) only 5%. For the material with 200–800 nm SiC added the amount of oxygen in gray matrix phase was impossible to determine using SEM even approximately due to the high density of Mg-B-O inclusions (figures 11a ,b). The observed oxygen segregation and "cleaning" of grain boundaries of MgB$_2$ from oxygen can be one of the reasons for the observed increase of critical current density critical current density (figures 10b, d) [66,69].

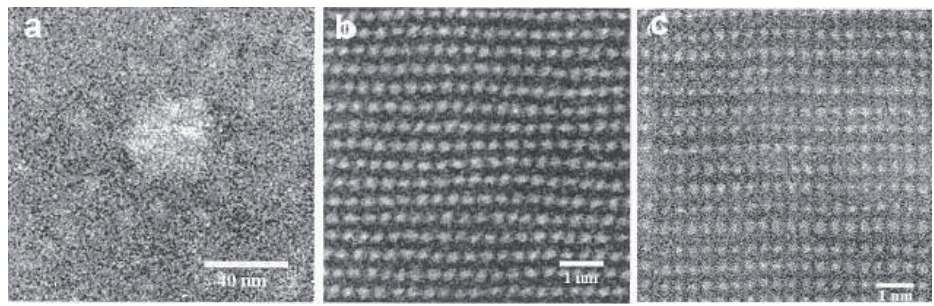

**Figure 7.** (a) Low magnification Z-contrast micrograph showing the brighter precipitate; (b) Z-contrast image of bulk MgB$_2$ in the [010] direction. The bright spots represent the Mg columns; pure B columns are not visible; (c) Z-contrast image of the coherent oxide precipitates in the bulk of MgB$_2$ [010]. A contrast variation in each second column is visible [68].

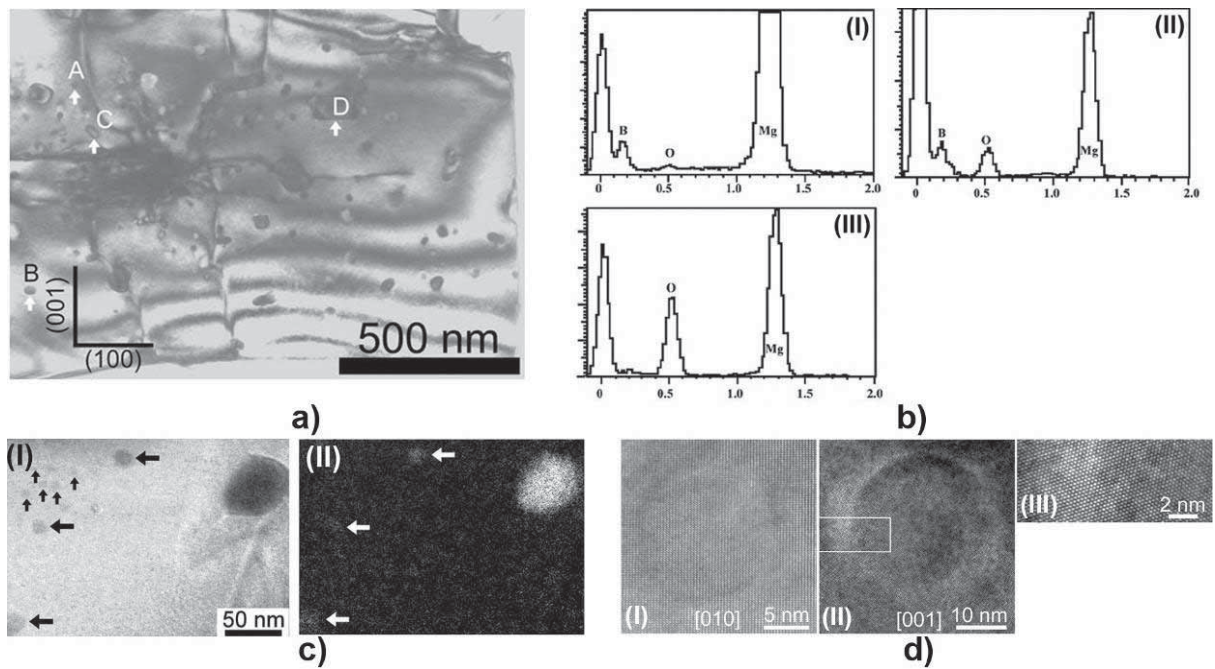

**Figure 8.** (a) [010] zone-axis bright-field diffraction contrast images of an MgB$_2$ crystallite. High density of precipitates of different shapes and sizes are clearly seen. Some precipitates are labeled with A, B, C, and D and indicated with arrows, demonstrating the change of shape with size and coalescence during the growth of the precipitates [68];
(b) EDX spectra of (I) the MgB$_2$ matrix, (II) a small precipitate, and (III) a large faceted precipitate [68];
(c) EFI images of (I) a boron map and (II) an oxygen map. Large arrows indicate precipitates with sizes ranging from 10 to 20 nm. Small arrows indicate precipitates about 5 nm in the size [68];
(d) HREM images of precipitates taken from (I) the [010] direction and (b) the [001] direction. The area marked with a white rectangle in (II) is enlarged and shown in (III) [68].

In the case of high-pressure (2 GPa) manufactured MgB$_2$-based materials some differences were found as compared to materials synthesized or sintered under the ambient pressure: (i) $J_c$ increases when SiC added if there is no notable interaction between SiC and MgB$_2$ (figure 12); (ii) Ti, Ta, Zr are usually bonded with hydrogen. The inhomogeneity of the oxygen distribution and formation of oxygen-enriched Mg-B-O inclusions in high-pressure synthesized materials were studied by scanning electron microscope ZEISS EVO 50XVP (resolution of 2 nm at 30 kV), equipped with: (1) an INCA 450 energy–dispersion analyzer of X-ray spectra (OXFORD, England), using which the quantitative analysis of elements from boron to uranium with a sensitivity of 0.1 wt % can be performed; probe 2 nm in diameter; (2) a HKL Canell 5 detector of backscattering electrons (OXFORD, England), which allows us to get (using the Kikuchi method) the diffraction reflections of electrons from 10 to 1000 nm areas and layers.

So, one of the mechanisms of an increase of $J_c$ in MgB$_2$ may be related to the formation of oxygen –enriched (Mg-B-O) inclusions, which can improve pinning, thus leading to an increase of the amount and length of grain boundaries (refinement of MgB$_2$ the structure), and to the "cleaning" of grain boundaries of MgB$_2$ from oxygen. The process of the Mg-B-O formation is sensitive to (i) synthesis or sintering temperature and usually its intensity increases with temperature; (ii) additions of Ti and Ta and, obviously, SiC can enhance the process of segregation.

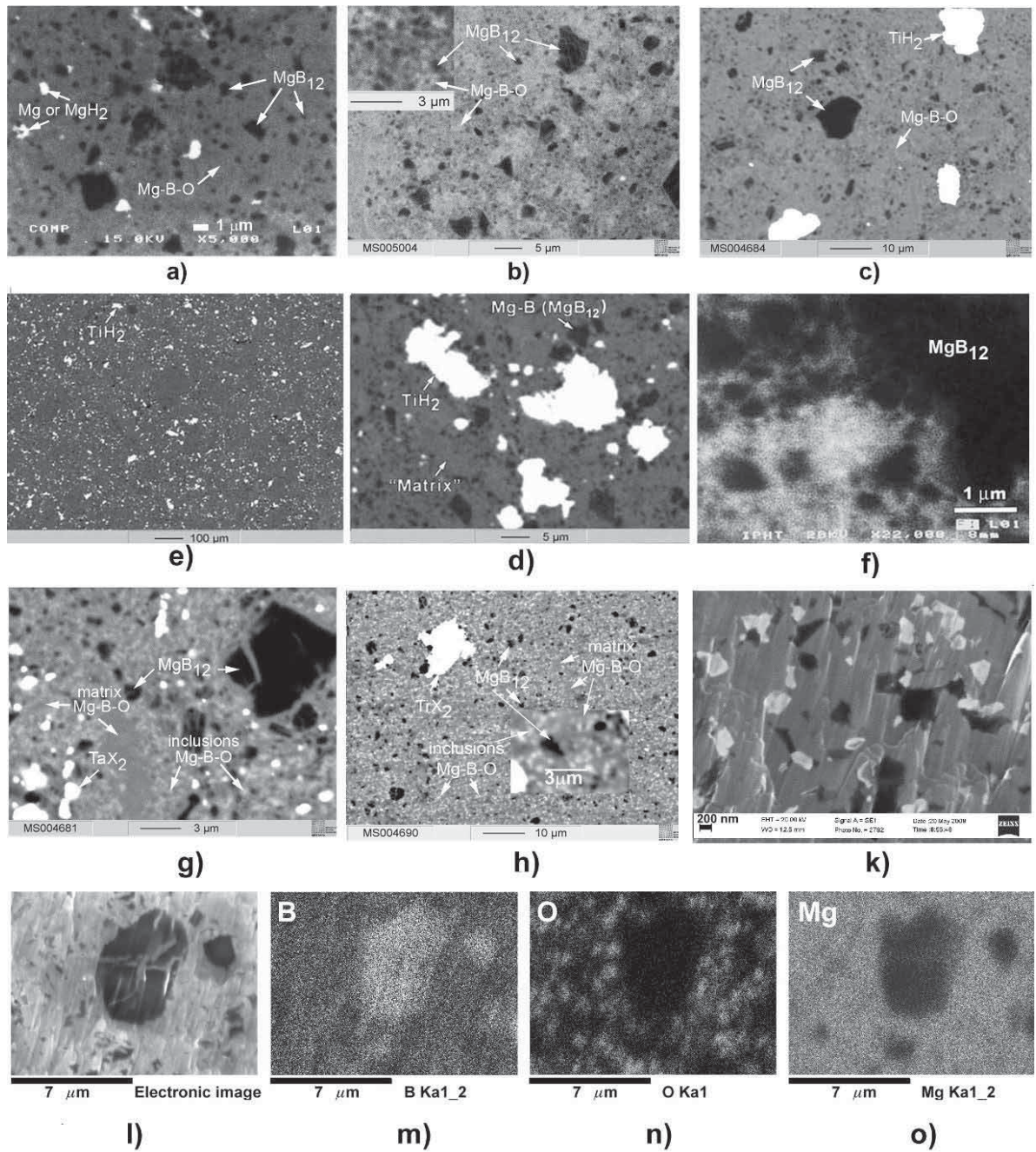

**Figure 9.** (a-i) Composition images (backscattering electron images) of high-pressure synthesized MgB$_2$-based materials at 2 GPa for 1 h from magnesium chips and amorphous boron Mg:B =1:2:
(a, e-f) at 800 °C from B MaTecK 95-97% purity, 0.8 μm, 1.7 % of O, (a) -without additions and (e-f)- with additions of 10 wt% of Ti under different magnifications;
(c, g-i) from B 4 μm, 1.5 % O: (c) at 800 °C with 10 wt% of Ti 1-3μm ; (g) at 1050 °C with 10 wt% of Ta 1-3μm;  (h-i) at 1050 °C with 10 wt% of Ti c-3μm under different magnifications;
(m-o) - distribution of boron, oxygen and magnesium, respectively, in the image shown in 8l (the place where Ti-containing phase is absent), the same sample is shown in 8h, k under different magnifications [41].

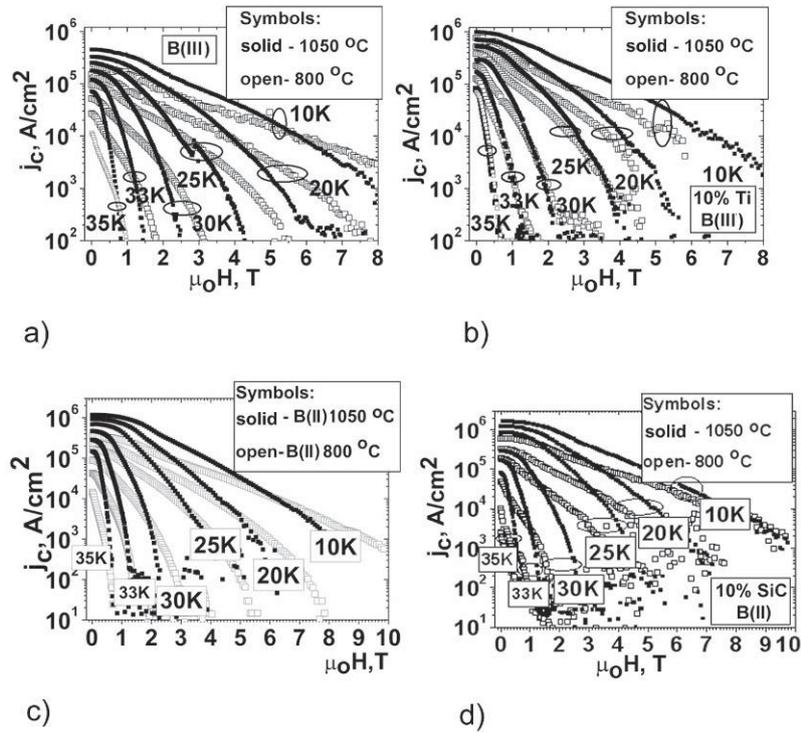

**Figure 10.** Dependences of critical current density (magnetic), $j_c$, at different temperatures on magnetic field, $\mu_0 H$, for materials synthesized from magnesium chips and amorphous Mg:B=1:2 at 2 GPa for 1h (a, b) from B 4 µm, 1.5 % O without additions and with 10 wt % Ti (1–3 µm), respectively (c, d) from B <5 µm, 0.66 % O without additions and with 10 wt % of SiC (200–800 nm), respectively. Solid symbols – synthesized at 1050 $^{\circ}$C, open symbols – at 800 $^{\circ}$C [66, 41].

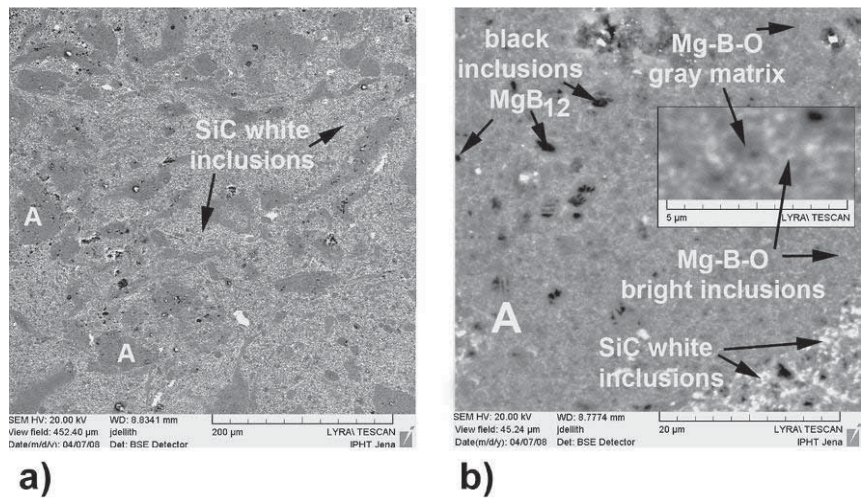

**Figure 11.** (a, b) Compositional images of MgB$_2$ samples synthesized from magnesium chips and amorphous boron (<5 µm 0.66 % O) Mg:B=1:2 with 10 wt% of SiC (200-800 nm) at 2GPa, 1050°C, 1 h under different magnifications. Figure 9b shows the typical structure of areas marked by "A" in figure 11a where SiC grains are practically absent [66].

Properties of MgB$_2$ bulk

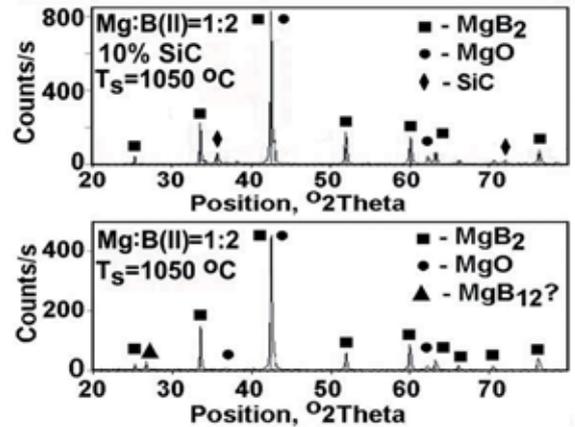

**Figure 12.** X-ray patterns of the materials synthesized from magnesium chips and amorphous boron (<5 μm, 0.66 % O) Mg:B=1:2 at 2 GPa, 1050 °C, for 1h: upper pattern - with 10 wt% of SiC (200-800 nm), bottom pattern – without additions.

*2.3. Higher borides and their influence on Jc.*

The influence of higher borides with near MgB$_{12}$ composition on superconducting properties of MgB$_2$ –based materials was studied in [40, 56, 66, 69, 70]. The effect of higher borides (seen in the SEM backscattering electron image as "black Mg-B inclusions") was observed earlier [38, 42, 71], but as it was already mentioned, the composition of these inclusions was miscounted (as single crystals of MgB$_2$ imbedded in Mg-B-O matrix). In [37] it was stated as well that the volume fraction of B-rich secondary phases was relevant microstructural parameters for the critical current density. As it was mentioned in [37] B-rich secondary phases are rarely addressed in the literature because it is hard to measure B accurately and determine the composition of these phases. It is hard or impossible to detect these phases by X-ray diffraction even if their amount is above 4% [72, 73]. The size and volume fraction of B-rich secondary phases (with near MgB$_{12}$ composition, in particular), is found to decrease with increasing synthesis, sintering or annealing temperature [40, 51, 56, 66, 69, 70, 74]. There still exists vagueness concerning appearance, composition and structure of higher boride phases: MgB$_4$, MgB$_6$ or MgB$_7$, MgB$_{12}$, MgB$_{16}$ or MgB$_{17}$, and MgB$_{20}$ [44, 45, 53, 54]. It is stated in [51] that equilibria of liquid phase with MgB$_{12}$ is metastable both at atmospheric pressure and at a pressure of 2GPa.

In the latest investigations [44, 45] the scheme of reaction between Mg and B at ambient pressure was presented as follows:

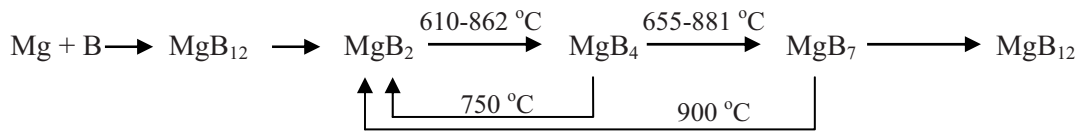

So, according to [44, 45], there exists two temperature intervals of the MgB$_{12}$ phase formation. The study of materials synthesis under high pressure (2–4 GPa) confirmed the low-temperature (around 800 °C) and high temperature (1000-1200 °C) intervals of the MgB$_{12}$ formation. The MgB$_4$ and MgB$_7$ were not determined at least up to 1100 °C in materials prepared from Mg:B=1:2 at 2 GPa and B-rich phase with near MgB$_6$ stoichiometry was present in materials synthesized under hot-pressing conditions (30 MPa) at 800-900 °C.

The notable effect of the presence of MgB$_{12}$ phase inclusions in MgB$_2$-based materials on SC properties compelled the authors of [40, 41, 66, 69, 75] to focus upon the synthesis of MgB$_{12}$ and B-rich phases contained in polycrystalline materials (figures 13, 14). From a mixture of Mg and B taken in MgB$_{12}$ stoichiometry a material has been synthesized at 4 GPa at 1000 °C for 1 h, which contained more than 50 % phase with near MgB$_{12}$ composition (marked D$_1$ and D$_2$ in figure 13a and appears black) and the rest of the sample contained magnesium oxide (segregated into large areas and appears white in figure 13a). But the most interesting fact is that the material turned out to be superconducting (figure 13 b). From mixtures of Mg and B taken in the 1:4, 1:6, 1:8, 1:10, 1:12, 1:17, 1:20 ratios materials with superconducting characteristics were synthesized at 2 GPa, 800 or 1200 °C for 1 h (figure 14d) [41, 66, 69]. The high critical current density and T$_c$ near 37 K was observed for the materials with near MgB$_{12}$ composition of the matrix (as the SEM and TEM observations show) prepared from the Mg:B =1:8 and 1:20 mixtures at 2 GPa, 1200 °C for 1 h. The structure and superconductive properties of material prepared from Mg:B=1:8 mixture is shown in figures 14 a–c, e, f. The corrected amount of shielding fraction (correction was made according to the sample shape and sizes) in the sample is 95.3%, which is indicative of a large volume of the SC phase in the sample. The sample (Fig. 14a) mainly contained a phase with near MgB$_{12}$ stoichiometry and some amount of unbonded MgB$_{(5.3-6.8)}$ grains as high-resolution TEM and SEM energy-dispersion analysis showed. MgB$_2$ phase was found by TEM as very random inclusions of size smaller than 100 nm (it was used JEM-2100F TEM equipped with an Oxford INCA energy detector, quantitative TEM-EDX analysis was performed using the Oxford INCA energy program, the diameter of probe being 0.7 nm). It should be mentioned that the materials with near MgB$_{16}$ stoichiometry of the matrix (synthesized, for example, from Mg:B=1:12 mixture at 2 GPa, 1400 °C, 1h) exhibited very pore SC properties.

On the basis of theoretical calculations metal-doped β-rhombohedral boron like LiB$_{13}$ [76] and MgB$_{20}$ [77] were believed to be high Tc superconductors. In [44, 45] it was stated that MgB$_{20}$ can be written as Mg$_{0.6}$B$_{12}$. The authors of [64], who synthesized an orthorhombic MgB$_{12}$ single crystal, space group Pnma with $a$ = 1.6.632(3) ˚ nm, $b$ = 1.7803(4) nm and $c$ = 1.0396(2) nm from the elements in a Mg/Cu melt at 1600 °C (Cu, Mg and B mixed in molar ratios 5:3:2), claimed that according to the preliminary results, the single crystalline samples were non-superconducting down to 2K. The results presented in [40, 41, 66, 69] show the probability of having SC properties for the MgB$_{12}$ compound. Unfortunately, the mosaic structure of grains with near MgB$_{12}$ stoichiometry (figure 14 b) gave no possibility for the authors of [41] to obtain sharp Kikuchi lines and to analyze their crystallographic structure. But the sharp T$_c$ transition, high critical current density (figures 14 e, f) of the material with the MgB$_{12}$ matrix as well as inconsistency of structural characteristics of MgB$_{12}$ reported in the literature (tetragonal or trigonal) still leave some hopes on high temperature superconductivity of the MgB$_{12}$ phase and point to the necessity of further investigations.

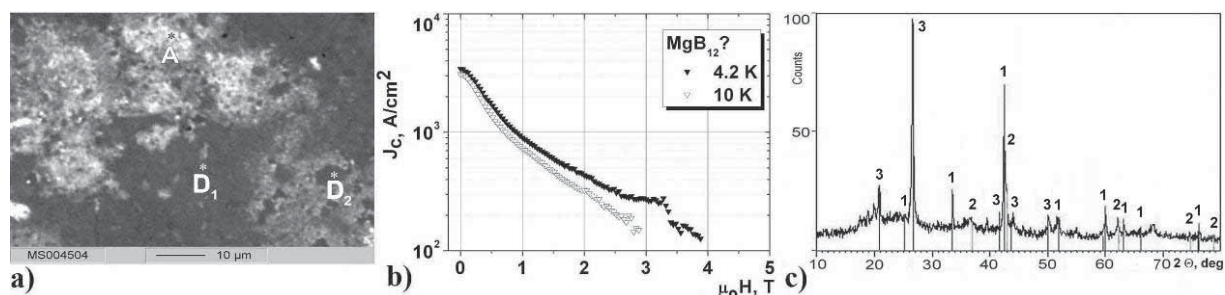

**Figure 13.** Structure (a), dependences of $j_c$ on $\mu_0 H$ (b) and X-ray pattern (c) of sample synthesized from mixed and milled by high speed planetary activator Mg chips and B (1.5 % O and 4 μm grains) taken in the MgB$_{12}$ stoichiometry at 4 GPa, 1000 °C for 1 h; in figure (a): A- bright region containing mainly MgO, D$_1$- 85% B, 13 % Mg, 1.5% O, D$_2$ - 86% B, 13 % Mg, 1.5% O. On X-ray pattern reflexes "1" and "2" coincide with those of MgB$_2$ and MgO, respectively, reflexes "3" possibly belong to MgB$_{12}$ or B-rich phase but reflexes "3" at 2Θ=27° and 2Θ=50° coincide with that of BN) [75].

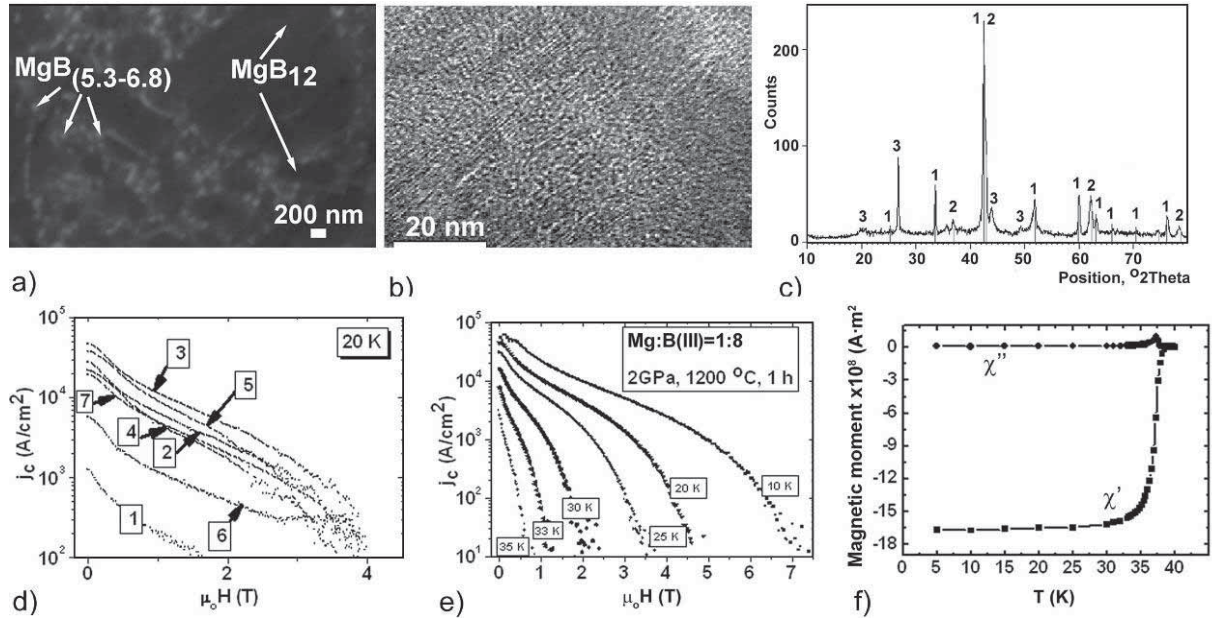

**Figure 14.** Structure and characteristics of materials prepared form mixed and milled by high speed planetary activator of magnesium chips and amorphous boron (4 μm, 1.5 % O) at 2 GPa, for 1 h:
(a) compositional image obtained by SEM ZEISS EVO 50XVP (resolution of 2 nm at 30 kV), equipped with an INCA 450 energy-dispersion analyzer of X-ray spectrums (OXFORD, England) and a HKL Canell 5 detector of backscattering electrons (OXFORD, England) of the sample synthesized at 1200 °C from Mg:B=1:8;
(b) TEM image obtained by JEM-2100F TEM equipped with an Oxford INCA energy detector of the grain with near MgB$_{12}$ stoichiometry in the material shown in figure12 a;
(c) X-ray pattern of sample shown in figure 12 a: (reflexes "1" and "2" coincide with those of MgB$_2$ and MgO, respectively, reflexes "3" possibly belong to MgB$_{12}$ or B-rich phase, reflex "3" at 2Θ=27° coincides with that of BN);
(d) dependences of $j$c on the external magnetic fields, μoH, at 20 K for the materials synthesized from Mg and B taken in the following ratio and synthesis temperatures, T$_S$: curves 1 – Mg:B=1:12, T$_S$ = 1200 °C, curve 2 – Mg:B =1:10 T$_S$ =1200 °C, curve 3 - Mg:B =1:8, T$_S$ =1200 °C, curve 4 - Mg:B =1:6, T$_S$ = 1200 °C, curve 5 - Mg:B =1:4, T$_S$ =1200 °C, curve 6 - Mg:B =1:12 , T$_S$=800 °C, curve 7- Mg:B =1:20, T$_S$=1200 °C;
(e, f) –characteristics of the material shown in figures 14 a-c synthesized from Mg:B= 1:8 at 2 GPa, 1200°C for 1h: $j$c vs μoH (e) and imaginary (χ'') and real (χ') part of the ac susceptibility (magnetic moment) vs temperature, T, measured in ac magnetic field with 30 μT amplitude which varied with a frequency of 33 Hz (f).

Irrespective of whether the MgB$_{12}$ phase is superconducting or not, its presence in the structure of MgB$_2$ – based material brings about an increase of critical current density (Table 3). Higher amount and higher dispersion of this phase (figure 2, for example) corresponded to higher critical current densities [40, 41, 56, 66, 69, 70]. The Ti and Ta additions to a MgB$_2$-based material affect the increased amount or volume fraction of the MgB$_{12}$ phase as it was shown for high-pressure synthesized material [38, 43, 56, 66]. The same effect can induce Zr. Inclusions of Ti- and Ta-containing phases (figures 9c-d, g, h) are rather coarse and too randomly distributed to effect pinning, but they absorb impurity hydrogen to form hydrides (figures 15 b, d, e). In the case of the zirconium borides formation (ZrB$_2$), figure 15 c, which occurred at higher synthesis temperatures, critical current density remained unchanged [41].

It is well known that boron and magnesium diboride react with moisture comparatively easy, and boron acids (H$_3$BO$_3$, HBO$_2$, H$_2$B$_4$O$_7$, etc.) can form. The formation of Mg(OH)$_2$ [47] and MgH$_2$ [38] in MgB$_2$ structure has been observed. Sometimes after synthesis material decomposes to form borane (hydrides of boron) having bad sharp smelt and one can even see how babbles evaporating on the surface of a sample. It is possible that borane can form during synthesis process and decrease the density of material or provoke the crack formation. The sources of hydrogen can be initial materials and materials with which MgB$_2$ is in contact during synthesis or sintering process. It was shown [43] that additions of TiH$_2$ (10 wt% ) instead of Ti resulted in a dramatic decrease of critical current density of high-pressure synthesized magnesium diboride and the material turns out to be rather porous (figure 16, material with TiH$_2$ addition with the best SC characteristics from synthesized in the 700-1000$^{\circ}$C temperature range is shown in this figure). Hydrogen can be solved in the MgB$_2$ structure. All these are cogent arguments for the fact that the presence of hydrogen in a material can harmfully affect its properties. The ability of Ti, Ta and Zr to absorb hydrogen and promote the formation of higher borides (with near the MgB$_{12}$ stoichiometry, in particular) allow one to increase $J$c and material mechanical properties due to the elimination of porosity and cracks formation, which can induce the presence of borane or hydrogen. The superconducting properties of a well high-pressure synthesized MgB$_2$-based material can stay unchanged for at least five years after storing in air.

**Table 3.** Critical current density, $j_c$, vs. relative amount, N, of "black inclusions" with near MgB$_{12}$ stoichiometry [66].

| Type of the initial B or MgB$_2$ | Manufacturing parameters: pressure, P, temperature, T, holding time, τ, | Name of the addition and its amount, wt.% | $j_c$ in 1T, at 20 K kA/cm$^2$ | N, % |
|---|---|---|---|---|
| MgB$_2$ (H.S. Starck, 0.8 % O, 10 µm) | P=2 GPa, T=1000 $^{\circ}$C, τ=1h | without | 5.2 | 1.8 |
| MgB$_2$ (Alfa Aesar, 98 % purity) | P=2 GPa, T=900 $^{\circ}$C, τ=1h | without | 7.8 | 2.6 |
| B (H.S. Starck, 1.9 % O, 1.4 µm) | P=2 GPa, T=800 $^{\circ}$C, τ=1h | without | 80 | 10.5 |
| B (H.S. Starck, 1.5 % O, 4 µm) | P=2 GPa, T=900 $^{\circ}$C, τ=1h | without | 62 | 9.5 |
| B (H.S. Starck, 1.5 % O, 4 µm) | P=2 GPa, T=1000 $^{\circ}$C, τ=1h | without | 240 | 12 |
| B (H.S. Starck, 1.5 % O, 4 µm) | P=2 GPa, T=800 $^{\circ}$C, τ=1h | without | 520 | 19 |
| B (MaTech ,95-97% purity, 1.6-1.7 % O, 0.8- 0.84 µm) | P=2 GPa, T=800 $^{\circ}$C, τ=1h | Ta, 2% | 90 | 10 |
| B (MaTech ,95-97% purity, 1.6-1.7 % O, 0.8- 0.84 µm) | P=2 GPa, T=800 $^{\circ}$C, τ=1h | Ta, 10% | 310 | 12.5 |
| B (MaTech ,95-97% purity, 1.6-1.7 % O, 0.8- 0.84 µm) | P=2 GPa, T=800 $^{\circ}$C, τ=1h | Ti, 2% | 95 | 10.7 |
| B (MaTech ,95-97% purity, 1.6-1.7 % O, 0.8- 0.84 µm) | P=2 GPa, T=800 $^{\circ}$C, τ=1h | Ti, 10% | 360 | 14 |

$^a$The amount of the "black" inclusions , N, was calculated as a ratio of the area occupied by the "black" inclusions at the COMPO image obtained at 1600x magnification to the total area of this image.

**Figure 15**. (a-e) X-ray patterns, dependences of $j$c on magnetic fields, $\mu_o$H, of the high-pressure synthesized MgB$_2$ –based materials from Mg chips and B (MaTech, 95-97% purity, 1.6–1.7 % O, 0.8-0.84 μm) taken in the Mg:B=1:2 ratio, mixed and milled in high-speed planetary activator with 10 wt. % additions of SiC (nanotubes 20–30 nm), Ti (1–3 μm) and Zr (2–5 μm). Regimes of synthesis and amount of additions are given in the pictures.

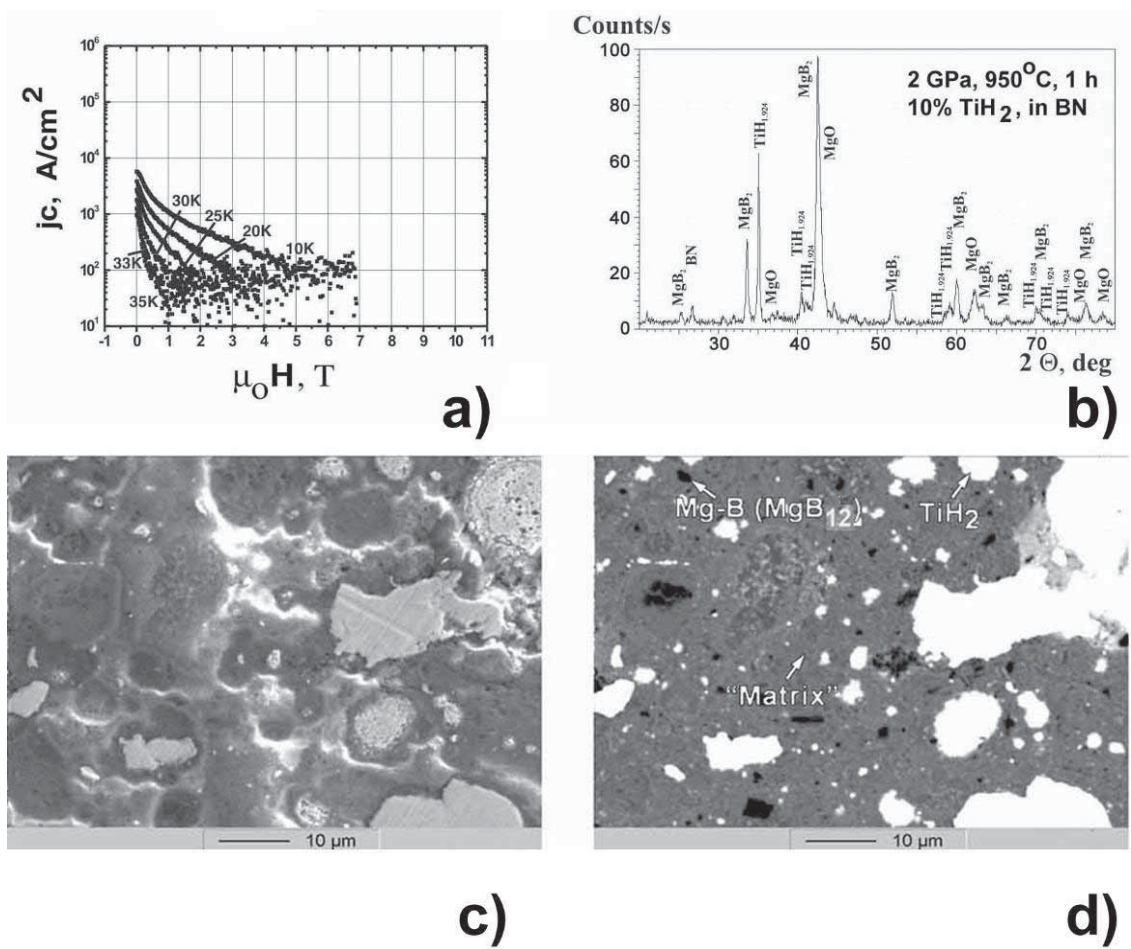

**Figure 16.** (a) Dependences of $J_c$ on magnetic fields, $\mu_o H$, of the high-pressure synthesized MgB$_2$ – based materials from Mg chips and B (MaTech ,95-97% purity, 1.6–1.7 % O, 0.8–0.84 μm) taken in the Mg:B=1:2 ratio, mixed and milled in a high-speed planetary activator with 10 wt % of additions of TiH$_2$; at 2 GPa, 950 °C, 1 h; (b) X-ray pattern of the sample; (c, d) structure of the sample obtained by SEM (c) SEI (secondary electron image) and (d) COMPO (backscattering electron image).

*2.4. Formation of titanium hydrides in the MgB$_2$ structure.*

The comprehensive study of the material with Ti addition high-pressure synthesized at 800 °C (figures 9i-f) indicates that the only titanium-containing compound in this material is a titanium hydride compound, TiH$_{1.924}$ [78]. The X-ray phase analysis, figure 15 b, shows the presence of only one Ti-containing phase TiH$_{1.924}$. EDX measurements support this result since the only detectable XRD elements present in the titanium-rich regions was titanium itself (hydrogen is not detectable by the EDX analysis). The presence of TiH$_{1.924}$ was surprising since titanium hydrides are relatively unstable compared to the more common titanium compounds, particularly the oxide phases for which one can expect the constituent elements to be readily available in these impure ceramics. Values of the formation enthalpy at the standard conditions illustrating the relatively low stability of titanium hydride are given in Table 4. Figure 17 (a) shows a bright–field TEM image of a particle from a finely ground sample of the high-pressure synthesized material (the only element detectable by EDX in this particle is titanium). A transmission electron diffraction pattern (TED) taken from this particle is shown in figure 17(b). Comparison with standard $d$-spacings from the JC-PDS database shows that the ratio of rings in this pattern corresponds closely to those predicted for titanium hydride TiH$_{1.924}$, and does not match very well any other Ti compounds in table 4. The microscope calibration has been attempted using diffraction patterns from adjacent MgB$_2$ grains as a standard, and the calculated $d$-spacing value for the first ring in the diffraction pattern of $2.58 \times 10^{-10}$ ($\pm 0.11 \times 10^{-10}$ m) is in very good agreement with the TiH$_{1.924}$ (001) $d$-spacing of $2.57 \times 10^{-10}$ m from the JC-PDS database.

Figures 17c-f show NanoSIMS ion maps (obtained by Cameca NanoSIMS 50 with a Cs+ primary ion beam) for the distribution of hydrogen, boron and magnesium ions and $^{48}$Ti$^{1}$H− ion clusters from a region of the MgB$_2$ matrix containing a number of titanium-rich particles. The detection of $^{48}$Ti$^{1}$H− ions demonstrates a close association between hydrogen and titanium in these particles. It is also clear that the yield of hydrogen is much greater from the titanium-rich particles than from the surrounding matrix. In order to decrease the possibility of the fact that this effect is caused by an enhanced yield of H from the vacuum system condensing on the surface of metallic Ti, the H yield from a bulk Ti metal standard under the same instrumental conditions has been analyzed. In this experiment the number of counts/second detected at unit mass is at least 100 times lower than from Ti-rich particles in the MgB$_2$ sample. These TEM and SIMS data very strong support the evidence of the presence of TiH$_{1.924}$ in these high – pressure synthesized materials. It has been suggested in [78] that hydrogen is a deleterious impurity in MgB$_2$, perhaps similar in behavior to oxygen presence of which can form weak links between MgB$_2$ grains [79]. It is not clear where the hydrogen in these high-pressure synthesized samples originates from: Mg, B and Ti source materials did not contain any impurity phases with hydrogen within the detection limit of XRD of about 1–2 vol.%.; possible hydrogen is being released from the apparatus or BN packing material during high-pressure synthesis or the surfaces of initial boron or MgB$_2$ powders may absorb some small amount of hydrogen which is not possible to detect by X-ray. The authors [78] consider that the reducing conditions created by the free hydrogen under the high-pressure conditions could explain why the much more stable Ti oxides at standard conditions compounds are not formed.

The same behavior demonstrated Ta and Zr. In the high-pressure synthesized materials prepared at higher synthesis temperatures 900–950 $^{\circ}$C, MgH$_2$ is usually absent in the structure of MgB$_2$–based materials, which indicates that hydrogen can escape from high-pressure apparatus at high temperatures. It turned out that as the synthesis temperature increases, Ti, Ta and Zr form borides. Sometimes borides and hydrides can form in parallel (see figures 9 g, h, where by "X" in compounds TaX$_2$ and TiX$_2$ boron or hydrogen is marked). It was said above that the formation of ZrB$_2$ at 950 $^{\circ}$C did not induce the $J$c increase (figure 15 c), while materials, in which ZrH$_2$ was present, demonstrated high absolute values of $J$c and increase in critical current density (as compared to undoped material, figures 15 b, d, e and 1b).

In [80-82] it was considered that positive effect of Ti and Zr on superconducting properties, and $J_c$ in particular, was associated with the refinement of MgB$_2$ structure. In [81] the positive effect of Ti or Zr on the increase of $J_c$ of MgB$_2$ has been explained by the formation of "the network " from TiB$_2$ or ZrB$_2$ grains, the size of which is about one unit cell. Such "network" refines the structure of MgB$_2$ and thus, leads to an increase in $J_c$. In high-pressure synthesized materials it is well seen that inclusions of

**Table 4.** Table comparing enthalpies of formation values for a range of common Ti compounds [78].

| Compound | Enthalpy of formation (kJ mol$^{-1}$) |
|---|---|
| Ti$_3$O$_5$ | −2459.4 |
| Ti$_2$O$_3$ | −1520.9 |
| TiO$_2$ | −944.057 |
| TiN | −336.6 |
| TiB$_2$ | −150 to −314 |
| TiH$_2$ | −15.0 |

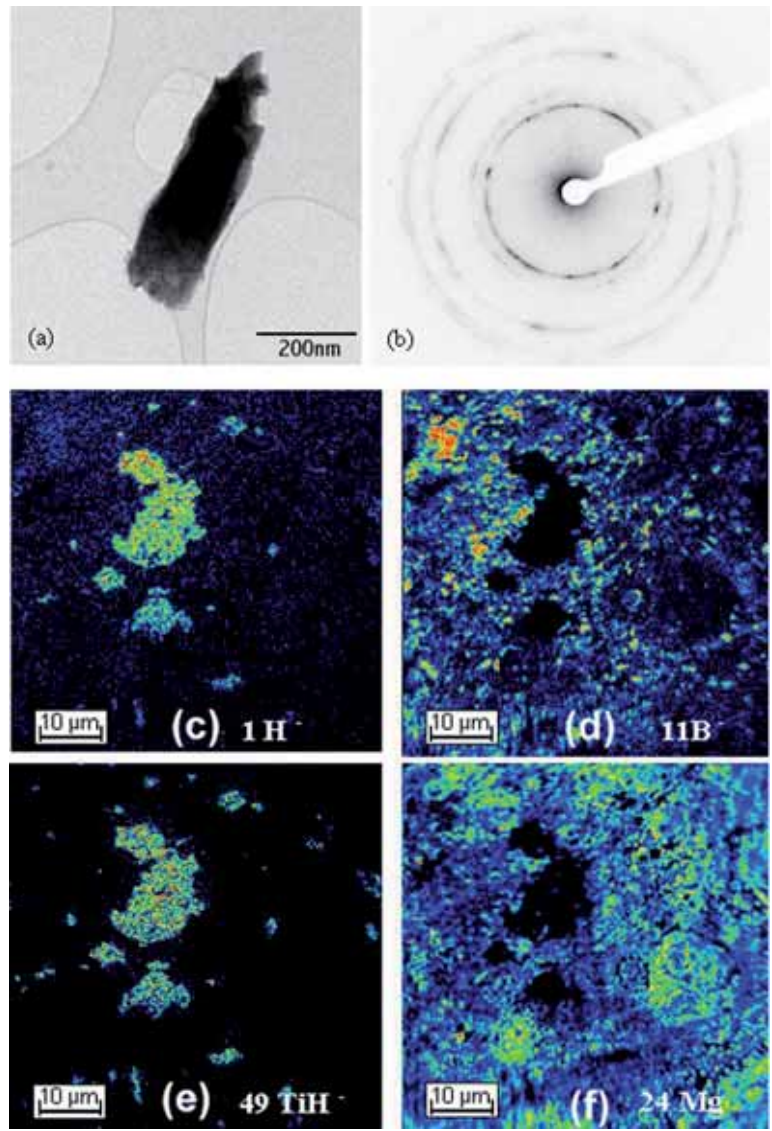

**Figure** 17. (a) Bright–field TEM image of a particle from a powdered sample of the high-pressure synthesized MgB$_2$ material with 10 vol.% Ti addition. (b) Electron diffraction pattern (contrast inverted) from the Ti-rich particle shown in (a). Transmission electron microscopy was performed on powdered samples dispersed on a lacy carbon film using a JEOL 3000F HR-TEM fitted with an energy dispersive X-ray (EDX) detector [78];
(c-f) NanoSIMS ion maps of the distribution of (c) mass 1 H−, (b) mass 11 B−, (d) mass 49 TiH$_2$−, (e) mass 24 Mg− (f) in high-pressure synthesized MgB$_2$ material with 10 wt% Ti addition. SIMS chemical analysis was carried out using a Cameca NanoSIMS 50 with a Cs+ primary ion beam. The particular importance of the NanoSIMS in this study is its ability to map the hydrogen distribution in these specimens.

Ti- or Zr- containing phase are rather coarse for being pinning centers and in addition are randomly distributed in MgB$_2$ structure. Most likely that the main effect can be due to the absorption admixture hydrogen, which in turn induces the redistribution of boron and oxygen to form B-enriched phases (MgB$_{12}$, in particular) at lower synthesis temperatures or oxygen-enriched Mg-B-O inclusions at high synthesis temperatures, whose sizes coincide with the coherence length of MgB$_2$, and very likely that they can improve pinning in MgB$_2$ (figures 9, 11).

## 3. Manufacturing of MgB$_2$-based material with high critical currents and upper critical fields and mechanical characteristics.

*3.1. High superconducting properties of bulk, wires and tapes MgB$_2$-based.*

Figures 18 show the dependences of critical current density, $J$c, on magnetic field, $\mu_oH$, estimated using vibrating sample magnetometer, figure 19 shows dependences of upper critical field, $H_{c2}$, on temperature, $T$, figure 20 shows dependences of field of irreversibility $H_{irr}$, as a function of temperature, $T$, figure 21 depicts superconducting transition temperature variation. There are highest superconducting characteristics observed for the today's possessing materials. Dozens of dopants have been found that can induce an increase of $J$c, $B_{c2}$ (or $H_{c2}$), $H_{irr}$, $T_c$ of MgB$_2$, which is detailed and comprehensively summarized, for example, in [4, 21]. The data shown in figures 18-21 have been obtained by different methods (magnetically or by direct current flow), in different laboratories using different devices and naturally, of course, with very different errors. Figures 18-21 reflect the properties of bulk materials, rods of wires and tapes (usually for the direction in which critical current density was the highest), tapes and wires in covering (or in shells) prepared using different methods and techniques from the initial materials of different purities and preliminarily treatment (mechanically activated, milled, etc.). It should be mentioned that it is practically impossible to control the amount of hydrogen and oxygen in the initial boron or MgB$_2$ powders, difficult to control the amount of carbon; besides, their amounts in the materials may dramatically vary with storing time and conditions. The same can be said about Mg. Because of the aforesaid it is rather difficult to compare properties of MgB$_2$-based materials and estimate the effect of dopants and mechanisms of their influence. The results given in figures 18-21 were intended to show the today's highest level of superconducting characteristics of MgB$_2$-based materials (with the exception of that for films) and to show that the characteristics of materials prepared from Mg and B or from MgB$_2$ without specially introduced additives (such as carbon, SiC, Ti, Zr, Ta, etc.) can be very high as well.

It may be even possible to conclude that each initial boron or magnesium diboride has its own "character" or "handwriting" and at present there are no methods to predict the optimal synthesis or sintering temperature on the basis of usually estimated by producer and given in certificates characteristics (even for the definite well-studied method of the MgB$_2$-based material synthesis or sintering), as well as it is impossible to predict without careful experimental study whether a definite dopant under definite manufacturing conditions will increase or decrease, for example, $J_c$ of the material: the same dopant (such as SiC, Ti, Ta, Zr) can essentially improve SC properties of a material prepared using one type of the initial boron or magnesium diboride and can dramatically reduce these properties if another type was used even if no notable difference in characteristics and purity of these types can be found. Besides, a dopant may increase $J_c$ of material at low temperatures, but decrease $J_c$ at higher ones or vice versa; may induce the increase of $J$c in high magnetic fields and reduce in the low one's or vice versa.

In general the higher critical current densities are usually more often observed for in *in-situ* prepared MgB$_2$-based materials than for the *ex-situ*, which one usually associates with finer MgB$_2$ grains in *in-situ* materials. For the doped materials (bulk, wires, and tapes) high $J$c values are reported

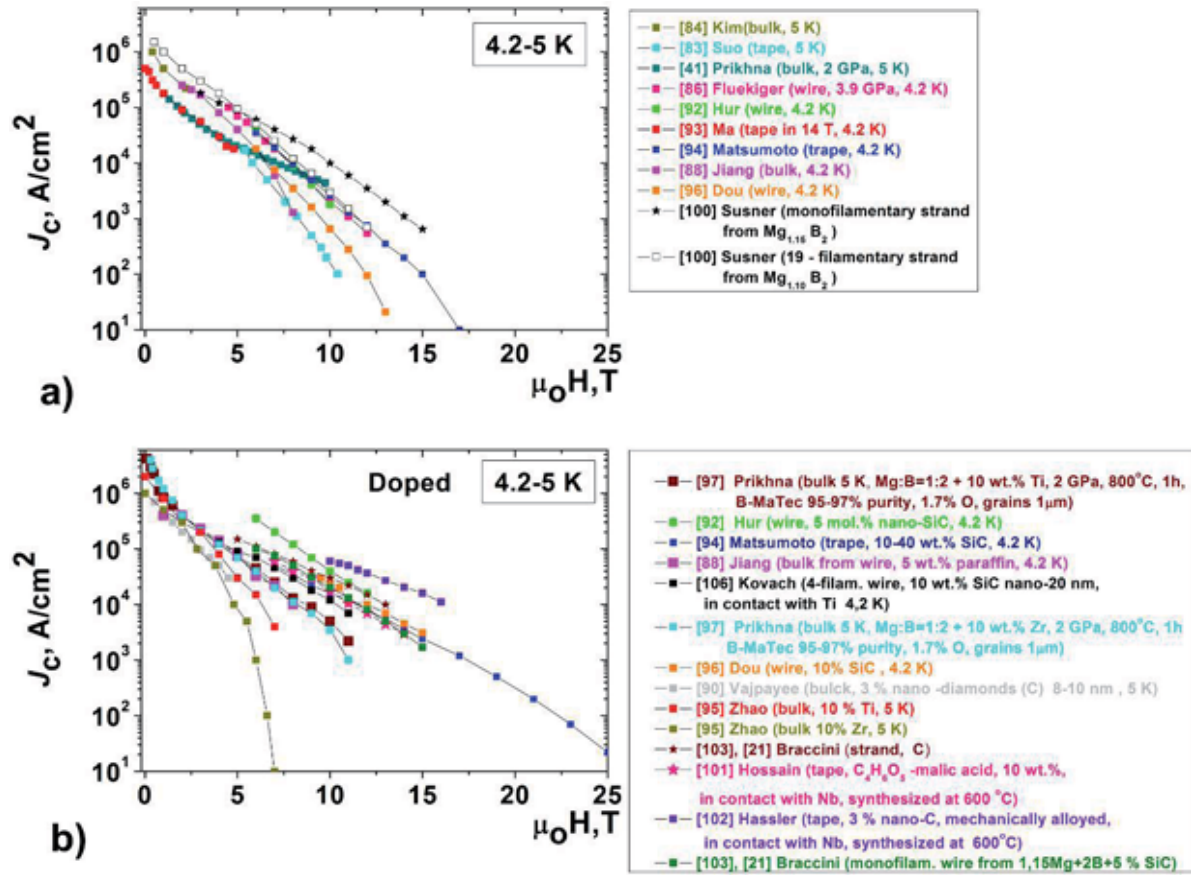

**Figure 18** (a,b) Dependences of critical current density (magnetic), $j_c$, at different temperatures on magnetic field, $\mu_o H$, for MgB2-base materials, at 4.2-5 K without (a) and with (b)dopants, respectively;

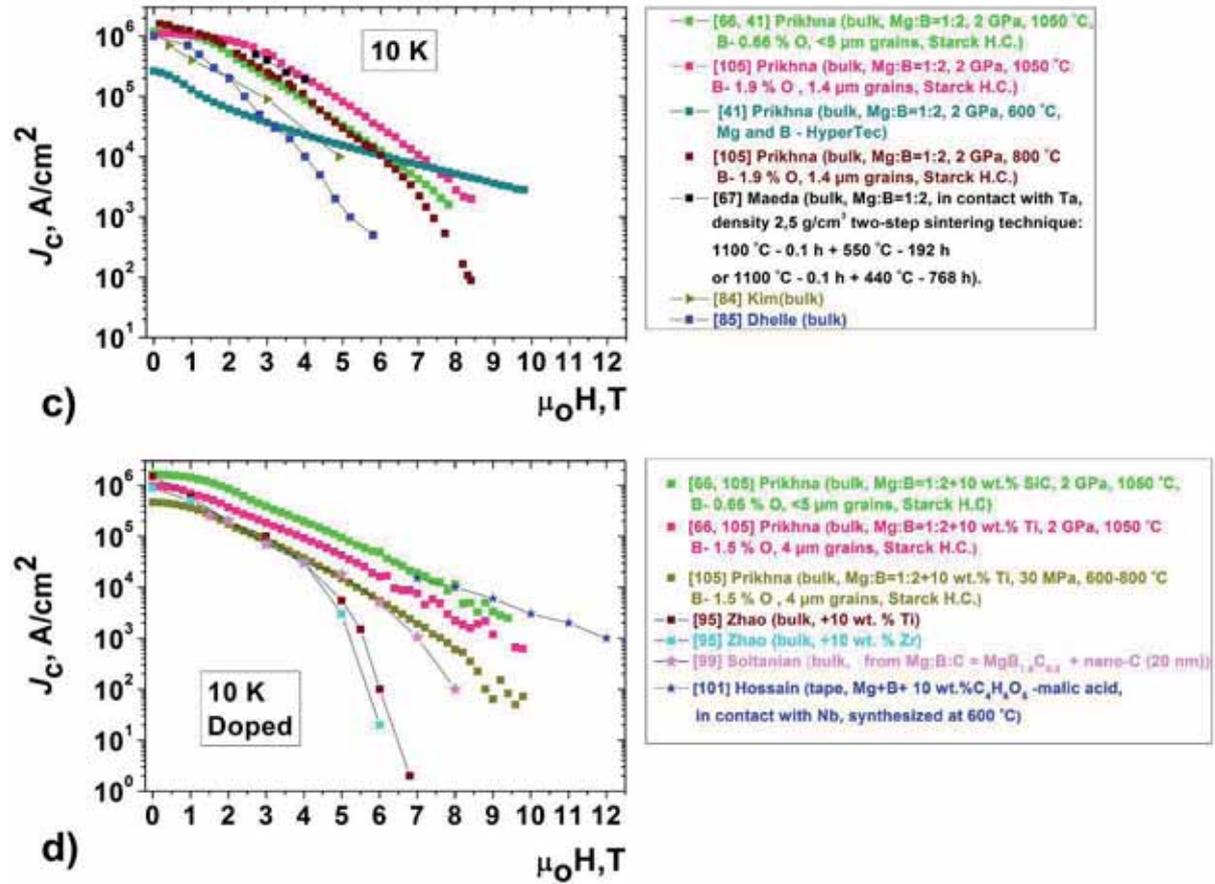

**Figure 18**. (Continuation) Dependences of critical current density (magnetic), $j_c$, at different temperatures on magnetic field, $\mu_oH$, for MgB$_2$-base materials (c, d) at 10 K without and with dopants, respectively;

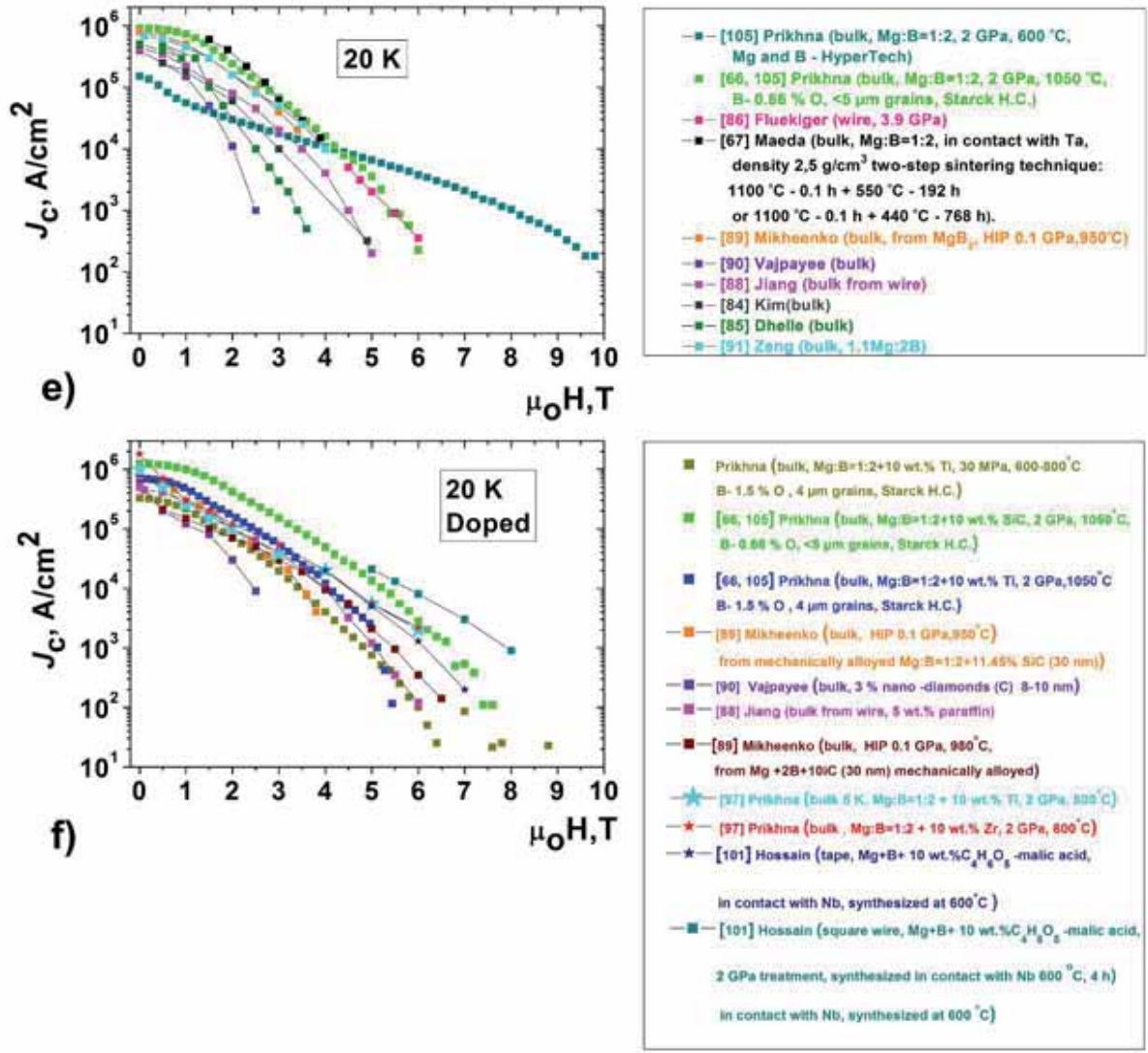

**Figure 18.** (Continuation) Dependences of critical current density (magnetic), $j_c$, at different temperatures on magnetic field, $\mu_o H$, for MgB$_2$-base materials (e,f) at 20 K without and with dopants, respectively;

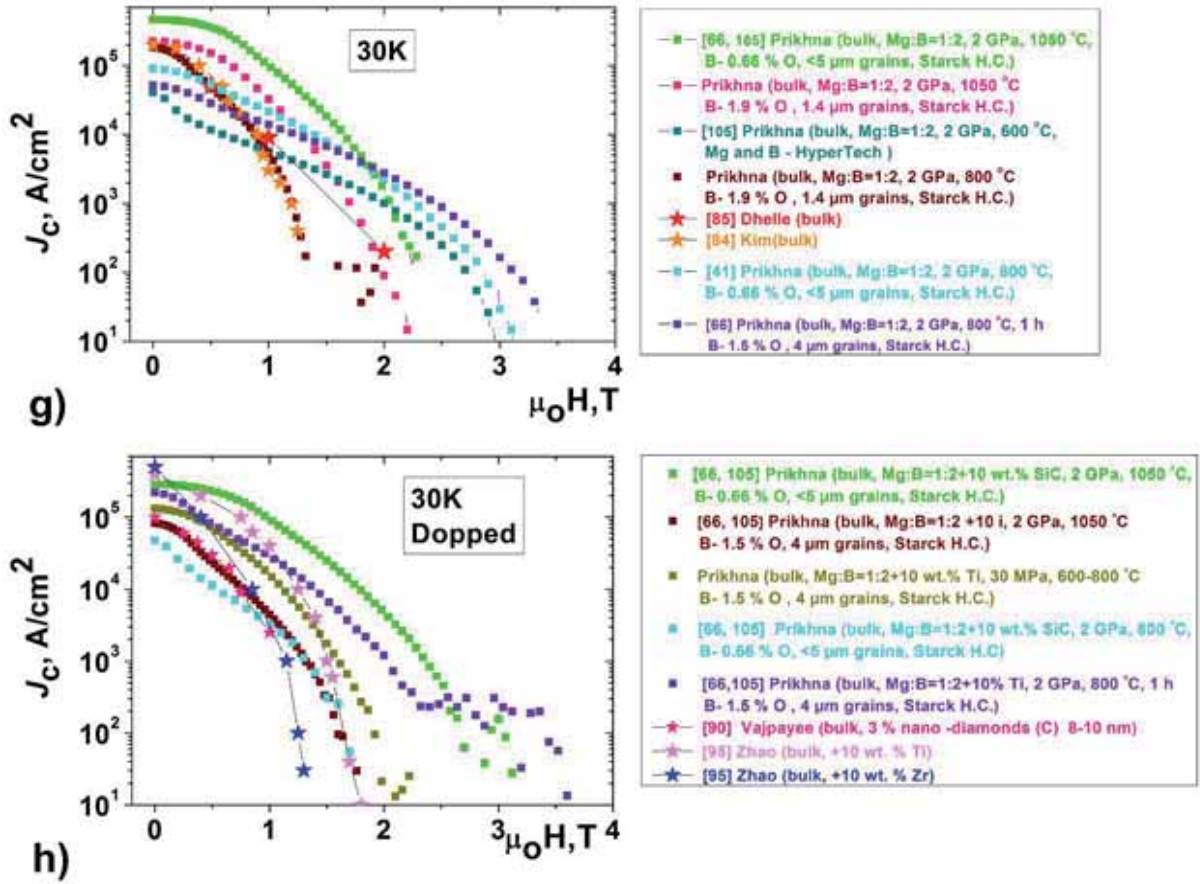

**Figure 18**. (Continuation) Dependences of critical current density (magnetic), $j_c$, at different temperatures on magnetic field, $\mu_o$H, for MgB$_2$-base materials (g, h) at 30 K without and with dopants, respectively;

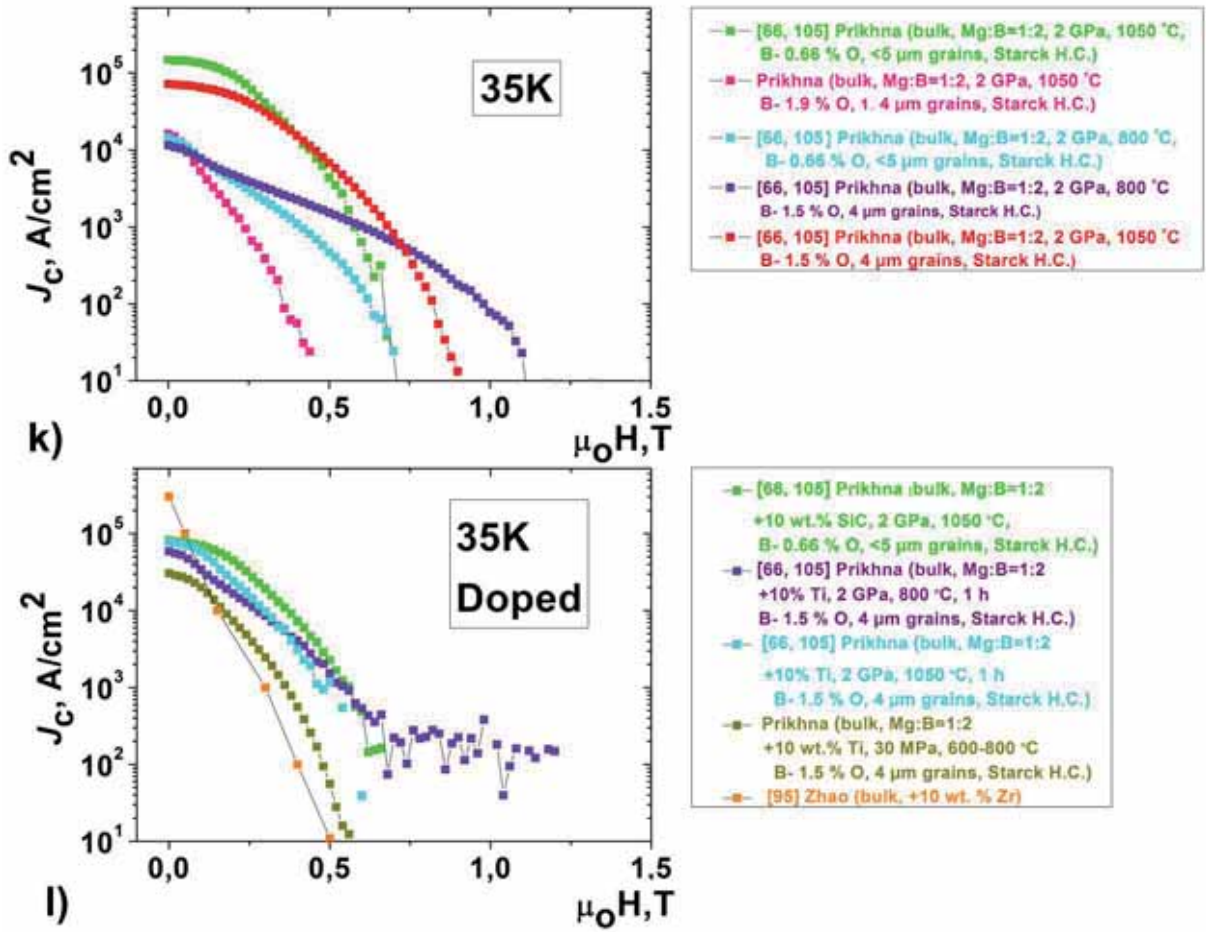

**Figure 18**. (Continuation) Dependences of critical current density (magnetic), $j_c$, at different temperatures on magnetic field, $\mu_o H$, for MgB$_2$-base materials (k, l) at 35 K without and with dopants, respectively;

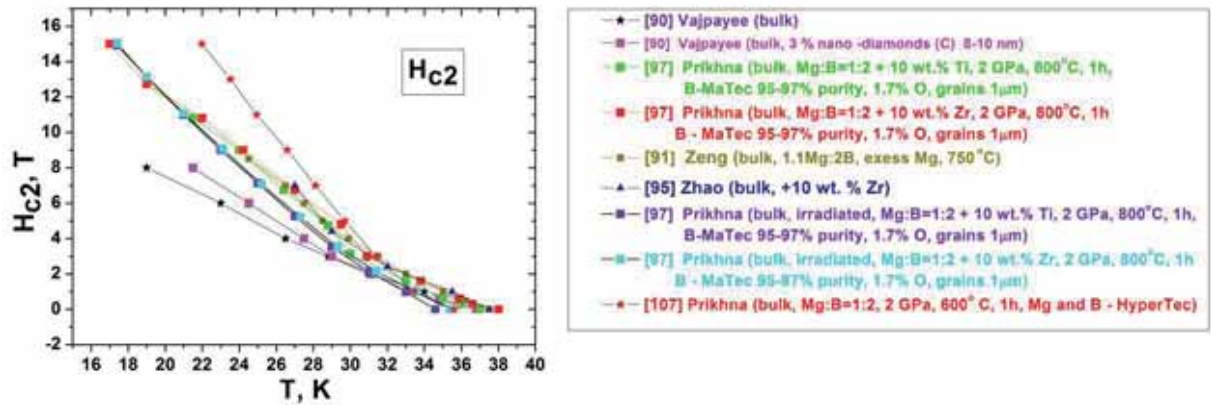

**Figure 19.** The upper critical field, H$_{c2}$, as a function of temperature, T.

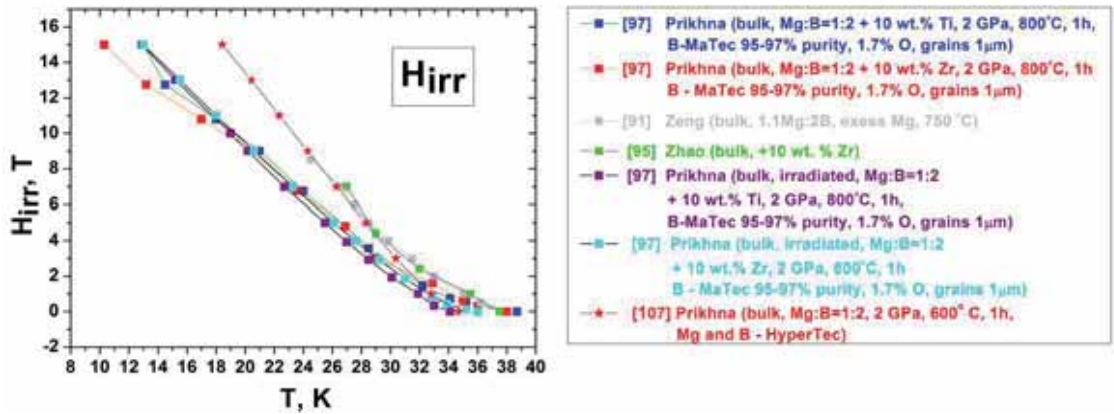

**Figure 20.** The field of irreversibility H$_{irr}$, as a function of temperature, T.

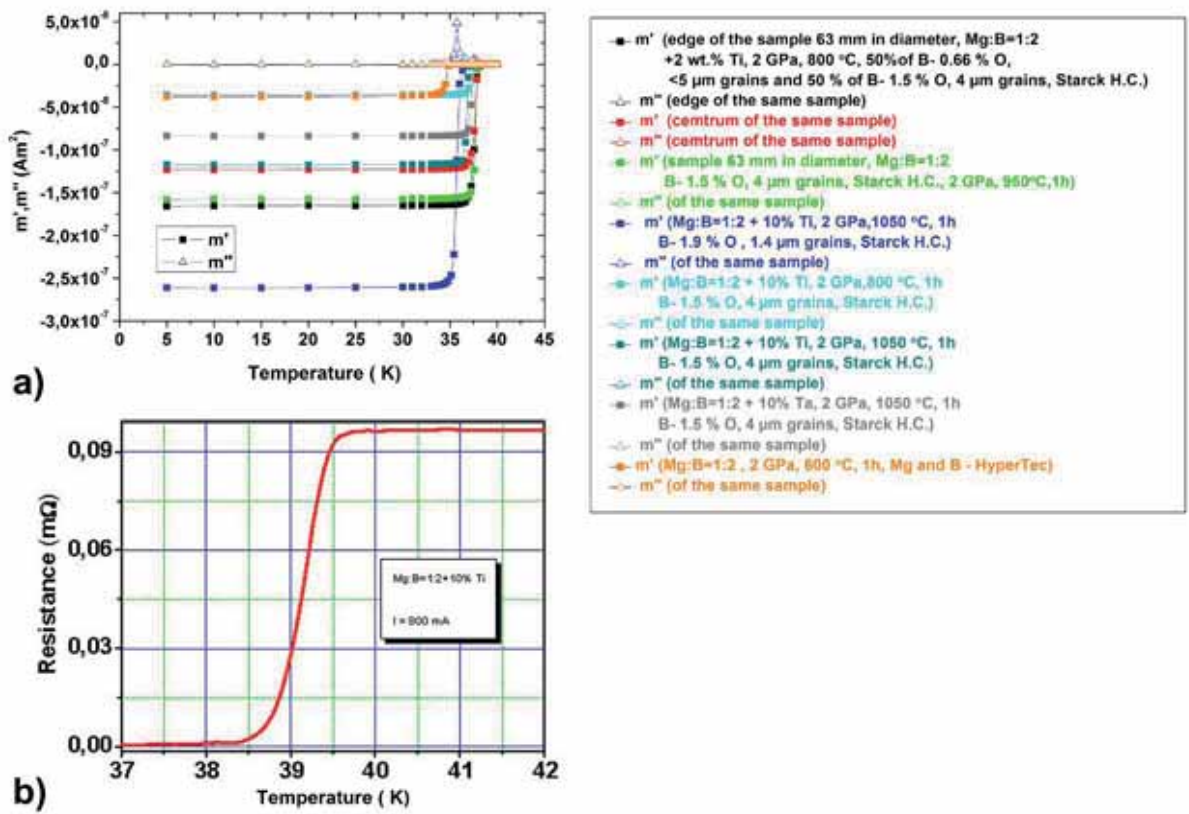

**Figure 21.** (a) Imagine (m') and real (m'') part of the ac susceptibility (magnetic moment) vs temperature, T, measured in ac magnetic field with 30 μT amplitude which varied with a frequency of 33 Hz (f) [107]; (b) SC transition for the material prepared from Mg:B=1:2 with 10wt. % of Ti (from amorphous MaTeck boron of 95-97 % purity) at 2 GPa, 800 °C, 1 h [107].

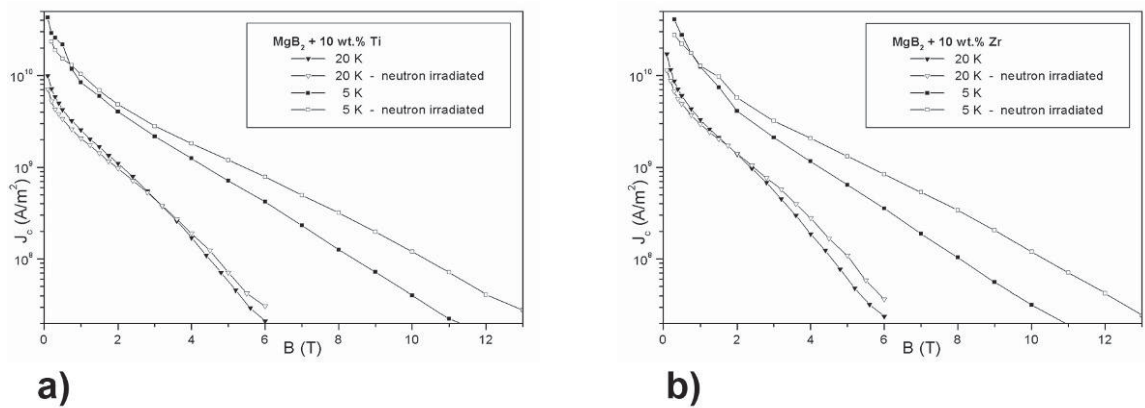

**Figure 22.** Critical current densities of the samples prepared from Mg:B=1:2 (amorphous B - 1.5 % O, 4 μm grains, Starck H.C.) at 2 GPa, 800 °C, 1 h with 10 wt.% of Ti or Zr addition before (a) and after (b) neutron irradiation at 5 K and 20 K [97].

for SiC doped, carbon (carbon nanotubes, nanodiamonds) or carbon sources (ethyltoluene, CH, sugar, maleic acid, maleic anhydride, paraffin, B$_4$C, toluene, acetone, ethanol, tartaric acid, etc.) doped and CaB$_6$-,Ti-, TiC-, Zr-, ZrB$_2$- doped materials [4, 21, 66, 96, 97, 108, 109]. Irradiation can only slightly increase SC properties of Ti and Zr-doped bulk high-pressure synthesized materials [97] and not over the whole range of investigated temperatures (figures 19, 20, 22). The $H_{c2}$ of Ti- and Zr-doped high-pressure synthesized material after irradiation was improved at low temperatures (figure 19) and the $J$c was improved at low temperatures and in high magnetic fields [97] (figure 22). A comparative study [96] of pure, SiC, and C doped MgB$_2$ wires has revealed that the SiC doping allowed the C substitution and MgB$_2$ formation to take place simultaneously at low temperatures. The C substitution enhances $H_{c2}$, while the defects, small grain size and nanoinclusions induced by C incorporation and low temperature processing are responsible for the improvement in $J$c. The irreversibility field ($H$irr) for the SiC doped sample prepared by the authors of [96] reached the benchmarking value of 10 T at 20 K, exceeding that of NbTi at 4.2 K. A systematic study on the effects of sintering temperature on the lattice parameters, C content, and electromagnetic properties allows the authors of [96] to demonstrate a unified mechanism, according to which the optimal doping effect can be achieved when the C substitution and MgB$_2$ formation take place at the same time at low temperatures (about 650 °C). According to [96], the C substitution is responsible for the enhancement in both $H_{c2}$ and flux pinning. C substitution for B induces a disorder in lattice sites, increases the resistivity and hence enhancement in $H_{c2}$, while C substitution together with low temperature processing results in a reduction of the grain size, fluctuation of $T_c$, extra defects, and embedded inclusions that enhance flux pinning. SiC doping takes advantage of both C substitution and low-temperature processing. An understanding of the dual reaction model has led to the discovery of the advantages of CH doping in MgB$_2$ by authors of [109], resulting in a significant enhancement in $J_c$, $H_{irr}$, and $H_{c2}$. CHs decompose at temperatures near that of the MgB$_2$ formation, thus producing highly reactive C, not dissimilar to the case of SiC doping. It is concluded in [96] that the proposed model has significant ramifications with respect to the fabrication of other carbon containing compounds and composites. But it was mentioned in [4] that an undoped and unirradiated seven-filamentary wire [110] demonstrated also a very high critical current density. A very high $J_c$ in the middle and high magnetic fields at 10–20 K exhibited bulk undoped material high-pressure (2 GPa) synthesized at 600 °C [105] (figures 18e, g). An essential improvement in $J_c$ can be attained in material produced from a mixture of Mg and B with excess Mg (as compared to MgB$_2$ stoichiometry) (figure 18e) [91]. As it was pointed in [91], the low-field $J_c$ for the 10% Mg excess samples sintered at 750 °C is increased by a factor of 3, compared to that for the normal MgB$_2$ samples, while the $H_{c2}$ for the 10% Mg excess samples sintered at 650 °C reached 8.7 T at 25 K, compared to 6.6 T for the normal sample. Besides, the Rietveld refinement X-ray diffraction (XRD) analysis showed that the MgO content was reduced in the 10% excess Mg samples, leading to an increase in the effective cross section of the superconductor [91].

The highest $J$c at low temperatures in magnetic field 10–16 T was reported for 4-filamentary wire prepared with addition of SiC in contact with Ti. Very often wires and tapes are prepared in contact with Ti or Nb layer, which is located between MgB$_2$ and steel. Intermediate layer from such a material can be a good getter of hydrogen and may be it is one of the reasons why using magnesium hydride as starting material, wires with high SC characteristics can be manufactured. In the earlier investigations of synthesis and sintering process under high pressure [42] the higher $J_c$ for the materials prepared in contact with Ta foil have been observed. The careful study of the possibility of reaction between Mg or B and Ta under the synthesis or sintering conditions has not shown any interdiffusion between these elements and only the transformation of Ta into Ta$_2$H (Ta is as well getter of hydrogen). Then the investigation of the influence of Ta adding to the initial mixture of Mg and B on high-pressure synthesized MgB$_2$-based materials along with other additives (Ti, Zr), which can absorb hydrogen, confirmed this effect [78, 66].

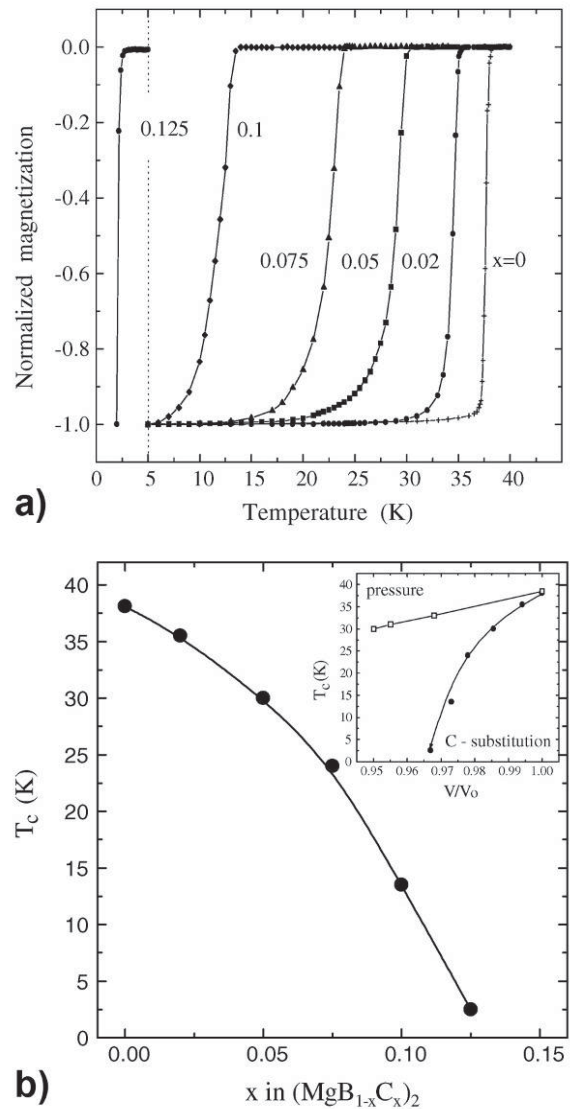

**Figure 23.** (a) Normalized zero-field magnetization measured at 10 Oe for the randomly oriented Mg(B$_{1-x}$C$_x$)$_2$ single crystals with a different C content x. (b) T$_c$ value determined by onset of magnetization vs nominal C content x in Mg(B$_{1-x}$C$_x$)$_2$. Inset show dependence of the T$_c$ on relative volume for C-substituted crystals in comparison to the data obtained for pristine MgB$_2$ (crystal B) produced in the hydrostatic pressure experiment [112, 113].

The data presented in figures 15 a and 10c, d (look as well figures 1a, b and 12) and the results described in [66, 111] demonstrate how differently SiC doping can influence high-pressure synthesized materials. In some contradiction with generally accepted opinion that nanosized (20–30 nm) SiC additions can be the most effective for an increase in $J_c$ and one of the reasons for such an effect is the interaction between MgB$_2$ and SiC, which results in incorporation of C into the MgB$_2$ structure, are the results for a high-pressure synthesized MgB$_2$-based material. The $Jc$ of high-pressure manufactured material with nanoSiC addition was slightly improved at 10 K in magnetic fields higher than 7 T, and at 20K in the fields higher than 5 T (figure 15a and 1a, b), but in medium and lower fields at 10 and 20 K $J_c$ stayed unchanged and at high temperatures (higher than 30 K) decreased essentially (the nano-SiC addition dramatically decreased $Jc$ of material synthesized at 800 $^o$C from the same initial boron). SiC with 200–400 nm grains slightly decreased the material $J_c$ over a whole range of temperature and field variations [111]. Only in the case of SiC with 200–800 nm grains added (figures 10d) when the interaction between SiC and MgB$_2$ (figure 12) was not notable (by X-ray) the $J_c$ of material was essentially improved practically over the whole range of temperature and field variations (with an exception of values in magnetic fields lower than 1 T at 30-35 K). A reaction between SiC, Mg and B under high-pressure and high-temperature conditions develops rather headily and it seems that too large amount of carbon is incorporated into MgB$_2$ lattice during synthesis time if SiC grains are small. It was shown by the authors of [112] (figure 23) that even with an insignificant increase of carbon content of MgB$_2$ single crystals Tc drastically decreases. For example, Tc of single crystal with composition Mg(B$_{0,875}$C$_{0,125}$)$_2$ was only about 2 K (figure 23 b). Besides, admixture carbon can be present in initial boron. Possibly the amount of carbon in dispersed boron may increase due to an absorption of atmospheric CO$_2$ during material storage. Additions of nanodimond (10 wt%) induced a dramatic decrease of $Jc$ of high-pressure (2 GPa) synthesized materials and rather essential shift and broadening of X-ray reflexes belonging to MgB$_2$ were observed. The low synthesis temperature proposed in [97] in the case of carbon doping with high probability limits the amount of carbon incorporated into the MgB$_2$ structure.

From the point of view of SC characteristics undoped MgB$_2$-based materials demonstrate high values, of $Jc$ in particular. The highest $J_c$ is attained for the materials with SiC, Ti, Zr or carbon-containing substance (toluene, maleic acid) doping. While in high magnetic fields (8T and higher) at 20 K the highest $J_c$ is demonstrated by high-pressure synthesized material without any dopants. In the case of carbon-containing dopants it is important for high $J_c$ that only definite comparatively small amount of carbon will be introduced into the MgB$_2$ structure. The positive effect of additions (such as Ti, Zr, SiC) can be as well due to the "refinement" of MgB$_2$ from hydrogen (by absorption) and from oxygen (due to the affecting its segregation) and due to the promotion of the formation of dispersed inclusions of higher borides. Each initial boron or magnesium diboride should be experimentally checked. Unfortunately, certificated characteristics of the initial materials provided are insufficient for predicting optimal manufacturing conditions for MgB$_2$-based materials with high $J_c$. Many variations of SC properties can be observed, especially if boron or initial MgB$_2$ is changed.

### 3.2 Mechanical characteristics of MgB$_2$-based materials

Up to now there are only several publications that describe mechanical properties of MgB$_2$ bulk materials. But for practical application in magnetic fields and/or in a termocycling regime the mechanical characteristics are of great importance.

Properties of MgB$_2$ bulk

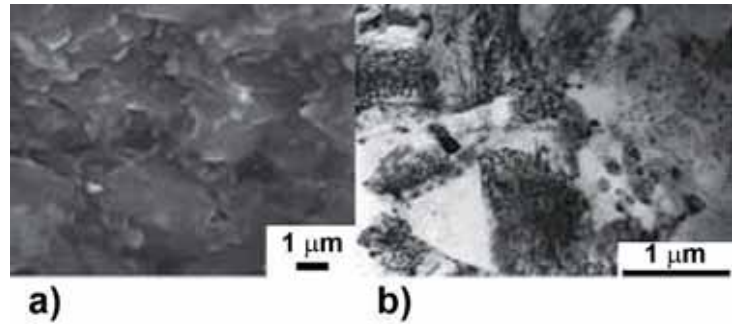

**Figure 24.** SEM micrograph and TEM bright–field image of HIPed sample [104].

In [104] it was reported about mechanical properties of bulk magnesium diboride prepared by high isostatic pressing from MgB$_2$ powder (*ex situ*) using the DMCUP HIPing cycle at 1000° C for the material with 2.666 g/cm$^3$ density, which was slightly higher than that of the theoretical one (2.625 g/cm$^3$) and which was indicative of the presence of admixture phases with higher density. The microstructure of typical studied material shown in figure 24 looks not completely dense but exhibited high $J_c$ in low magnetic fields (at 10 K about 10$^6$ A/cm$^2$ in zero field and 2·10$^4$ A/cm$^2$ in 5 T field; at 20 K about 8·10$^5$ A/cm$^2$ in zero magnetic field and 4.6 T field of irreversibility).

The average Vickers microhardness measured for this material under a load of 1 kgf was equal to 10.4 GPa for the sample prepared from non-milled powder and 11.7 GPa for the sample from milled powder [104]. The highest values of the Young ($E$) and bulk ($K$) moduli and Poisson ratio ($v$) calculated based on the elasticity theory were: E=272.5 GPa, K=142.5 GPa, $v$= 0.181. Fracture toughness $K_{1C}$ was found to be 2.17 MPa·m$^{1/2}$ and 2.41 MPa·m$^{1/2}$ for samples HIPed from non-milled and milled powders, respectively. These values of fracture toughness are between the values for single–crystal silicon and single–crystal Al$_2$O$_3$ (sapphire) [104].

In the case of high-pressure synthesized materials [42, 114], mechanical characteristics (micro-, macro- and nanohargness, fracture toughness, Young modulus) were investigated as follows:
1) Macrohardness was determined using "ТП-2" hardness meter equipped by Vickers indenter.
2) The Vickers microhardness was determined using microhardness meter "ПМТ-3" (ЛОМО, Russia) under the load from 1 up to 10 N. The sizes of indents were measured by universal exploratory NU-2 (Carl Zeiss Jena, Germany) under 750 magnification in the regime of phase contrast. The microhardness was calculated using equation (3):

$$H_V = 0{,}4636\, P/a^2, \qquad (3)$$

where P is the indentation load, a is the half of indent diagonal (figure 25).

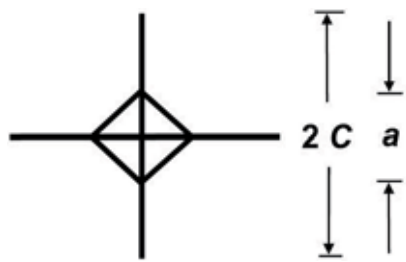

**Figure 25.** The scheme of indent measurement.

3) Fracture toughness, $K_{IC}$, which characterizes the material resistance to crack propagation and is the important mechanical property of brittle materials to which MgB$_2$-based superconductive ceramics belongs, was investigated using the indentation method of fracture toughness estimation based on the phenomenon of crack formation when local loading of brittle materials is performed. For the Vickers indent Evans and Charles [115] have derived the equation

$$K_{IC} \cdot \Phi/Ha^{0.5} = 0.15k(C/a)^{-1.5} \qquad (4)$$

where $\Phi$ is the constant of Marsh ($\approx 3$), H – Vickers hardness, a – half of the indent diagonal, C – average length of the radial cracks measured from the center of indent (figure 25) k = 3,2.
The value of k was determined empirically, using $K_{IC}$ values measured by the standard methods at macroscopic samples. Using the Vickers hardness (3), equation (4) can be reduced to the following:

$$K_{IC} = 7.42 \cdot 10^{-2} \, P/C^{1.5} \qquad (5)$$

4) Nanohardness and Young modulus was determined on a Nano Indenter II nanohardness tester (MTS System Inc., USA) using diamond indenter which has the shape of triangular pyramid (Berkovich indenter). Using this method it is possible to determine mentioned above characteristics even of a single grain (if it is about 10-6 μm in size). The device allows the measurement of the hardness in the range of loads from 0.001 g up to 15 g ($10^{-5}$- 0, 15 N). The minimal depth of indent that is enough for measurement of the hardness and Young modulus is $\approx$30 nm. The main working regime is the loading with a constant speed of its increasing; the load can be varied from 0,0001 to 20 mN/s. The accuracy of the indentation is ± 400 nm. The device allows one to measure the unreduced hardness under the load and the Young modulus according to the unloading curve (the size of the indent under the given load smaller that the grain size or equal to it.

The microhardness of the high pressure (2 GPa) –high temperature (800 °C) sintered material estimated using a Vickers indenter ($H_v$) under a load of 4,9 N was: 12.65±1.39 GPa (for the material without additions), and 13.08±1.07 GPa with 10% Ta added and 12.1 ±0.08 GPa with 2% Ti added. The Vickers hardness under a 148.8 N-load was $H_v$=10.12±0.2 GPa. Because of the absence of cracks under a load of 4.9 N, the fracture toughness was estimated at a load of 148.8 N: $K_{1C}$=4.4± 0.04 MPa·m$^{0.5}$ for the material without additions and $K_{1C}$=7.6± 2.0 MPa·m$^{0.5}$ with for the material with addition of 10% Ta [42, 114].

The mechanical properties (Berkovich nanohardness and Young modulus) of MgB$_{12}$ inclusions (of about 10 μm in size) in MgB$_2$ matrix of high-pressure sintered material were estimated using Nano Indenter II nanohardness tester and presented in [42, 114, 66]. The nanohardness of inclusions with near MgB$_{12}$ stoiciometry at a 60 mN-load is 35.6±0.9 GPa, which is higher than that of sapphire (31.1±2.0 GPa), while nanohardness of the "matrix" under the same load is 17.4±1.1 GPa only. The Young moduli are 213±18 GPa of the "matrix", 385±14 GPa of the inclusion and 416± 22 GPa of sapphire.

The Vickers microhardness ($H_v$) of the material with near MgB$_{12}$ composition of matrix (figure 14 a) [66], prepared at 2 GPa, 1200 °C, 1 h was twice as high as that of MgB$_2$ (25±1.1 GPa and 12.1±0.8 GPa, respectively, at a load of 4.9 N.

Despite the fact that indentation load is included into the formulas for estimation of mechanical characteristics, in the case of ceramic materials for comparison of the data it is important to mention the indentation load because of the elastic aftereffect and changing of the indent shape after removing indenter from a ceramic material. The elastic aftereffect results in increasing of estimated values of microhardness and fracture toughness as the indentation load decreases. The hardness or microhardness (estimated under comparatively high load) is an integrated characteristic and gives a notion about mechanical properties of a material as a whole, because the area of the indent is rather high and many grains of material appear under the indent. Using the nanohardness one can characterize mechanical properties of separate phases, which are present in the material if their grains or occupied by this phase area is enough for performing the study.

The highest values of mechanical characteristics for bulk MgB$_2$ –based materials measured up to now are as follows: Vickers hardness under a 148.8 N-load is H$_v$=10.12±0.2 GPa and the fracture toughness under the same load is K$_{1C}$=7.6± 2.0 MPa m$^{0.5}$; Young modulus is E=273 GPa. Inclusions of higher borides with near MgB$_{12}$ composition appeared to be harder than sapphire and approximately two times harder than the MgB$_2$-based material matrix. The microhardness of the material with near MgB$_{12}$ composition of the matrix demonstrates an unusually high, as compared to MgB$_2$ microhardness, which one more time gives grounds to conclude that MgB$_{12}$ phase as well as sample with MgB$_{12}$ matrix, which structure is shown in figure 14 a are principally different from MgB$_2$.

## 4. MgB$_2$-based bulk materials for practical application

Superconducting materials bring several basic benefits to electric power. Zero dc resistance of superconductors offers significant improvements of the efficiency in generating, delivering, and using electric power. A secondary but equally important benefit of the low resistance is the possibility to attain a very high current density [116].

A considerable growth of the superconductive material market is expected to be due to the application of MgB$_2$-based material, in particular, operating at liquid hydrogen (20 K) or neon (27 K) temperatures. The possible practical applications of MgB$_2$ are similar to that of melt-textured YBa$_2$Cu$_3$O$_{7-\delta}$ (MT-YBCO). Such materials can be used in the fault current limiters, cryogenic electromotors and generators, for the purposes of shielding, in the pumps for liquid gases pumping, for manufacturing of magnetic bearings (for flywheel energy storage devices, for high-speed rotating centrifuges and wind mills, for example), as trapped-field magnets (in the NMR spectroscopy to create high magnetic fields, in undulator of a synchrotron storage ring, in magnetron sputtering device with superconducting trapped-field magnets installed as a magnetic source, in maglev transport, and in other cryogenic devices [116, 117].

According to the market investigations of Bento Strategy described in their report 'Superconductor market research HTS in 2009" distributed by e-mail, the fault current limiters can be one of the primary applications of high-temperature superconductors (HTS) in electric power systems. At the next step are current leads and NMR, big expectancies are given to electrical motors, generators and power storage devices. The main perspectives for bulk MgB$_2$-based material application are seeing in electrical devices such as superconducting electromotors (generators) and pumps, inductive fault current limiters and shielding screens.

The advantages of using superconductors in the above applications [116-118] are as follows: using superconducting (SC) elements in cryogenic motors (generators and pumps) allows one to make them much smaller and lighter, improve the efficiency and power factor, increase the response speed due to the lower inertia moment. The fault current limiters (FCLs), inductive FCLs, in particular, have no analogs in the traditional energetics. In contrast to other power superconducting devices, which are designed to increase efficiency of power equipment by means of their lower losses, smaller size and weight, the FCLs are designed to increase the reliability of power delivery and to protect electric power devices against the electrodynamic and thermal stresses.

The problem of fault currents is one of the oldest problems of electrical engineering. It is estimated that about 150 faults per year and per 100 km of line occur in the distribution lines. When a fault occurs in the medium or high voltage networks, the short-circuit current may increase to values as high as 30-60 times the rated current. The conventional circuit breakers interrupt these short-circuit currents after about 50-60 ms, special fast circuit breakers can do it faster. However, the principle of a circuit-breaker is to cut the fault current at its zero crossing, and hence, all the components of electric systems have to withstand the destructive effects (thermal and electrodynamic) of the fault currents for a period of at least 10 ms. Electric power system designers often face fault-current problems when expanding existing buses. Larger transformers result in higher fault-duty levels, forcing the replacement of existing buses and switchgear not rated for the new fault duty. The current-interrupting capability of circuit breakers often becomes insufficient when additional power sources are included into the network [118].

Interest for the fault current limiters and electric machines that are working at liquid hydrogen temperature caused by the modern technological progress aimed toward the technologies utilizing liquid hydrogen, for example, at the development of the electrical power networks, by which the electrical power should be transmitted for long distances through the superconductive cables at the liquid hydrogen temperature and using liquid hydrogen as cooling agent (liquid hydrogen also may be transmitted for long distances), at the use of liquid hydrogen as ecologically clean fuel for different types of transport, etc. The use of MgB$_2$, which has lower as compared to YBa$_2$Cu$_3$O$_{7-\delta}$ (93 K) superconducting transition temperature (39-40 K) can be explained by a comparatively simple and cheap manufacture of MgB$_2$ - based materials and by a possibility to get higher critical currents in polycrystalline material.

The advancement in applications of the superconductors in SC FCLs, electrical machines and other devices working on the principles of levitation and diamagnetism is closely related to the material improvement, because when operated they should provide high currents in magnetic fields, their structure should be highly homogeneous on nanostructural level and possess high mechanical characteristics for providing the high resistance to thermocycling and withstanding against stresses induced by magnetic fields. At present one of the most suitable materials for such purposes are polycristaline (nanostructural) MgB$_2$-based superconducting materials. But further improvements of their characteristics as well as the development of production technologies of components from these materials for cryogenic devices are of great importance.

MgB$_2$-based materials can be used for shielding (in cryostats, etc.). The manufacturing technologies developed allow us to produce dense mechanically stable and resistive to degradation (from the point of view of $J_c$ stable for five years stay in air) materials and large components with high critical current densities. The price for high-pressure synthesized magnesium diboride is about 600-900 EUR per kilogram (it is about 3 times cheaper than MT-YBCO material). High superconducting characteristics can be realized in the polycrystalline structure (nanostructure) and at present items in diameters up to 150 mm can be produced. Hot pressure technique (25-30 MPa pressures) makes it possible to manufacture MgB$_2$-based bulk materials with high superconducting (figure 26c) and mechanical properties (for example, in the shape of a cylinder up to 250 mm in diameter). At present there are practically no limits in producing items by hot pressing of about meter in diameter.

The encouraging results from the point of view of the material density and SC characteristics have been achieved using spark plasma sintering (SPS) for *ex situ* and *in situ* preparation of MgB$_2$ [119-125]. This recently developed SPS method has features of both high-pressure and hot pressing sintering and allows producing bulk near theoretically dense rather large ingots and parts. The transition temperature higher than 38.5 K and a superconducting transition width less than 0.5 K for *ex-situ* SPS-prepared MgB$_2$ have been registered in [125]. In addition, high critical current densities of $J_c = 7.7 \cdot 10^5$ A/cm$^2$ in a field of 0.6 T at 5 K and of $8.3 \cdot 10^4$ A/cm$^2$ in a field of 0.09 T at 35 K were observed. The comparatively high superconducting characteristics of the SPS-processed MgB$_2$ were attributed to the presence of MgO nanoparticles and to the uniformly distributed dislocations inside the MgB$_2$ grains.

In principle it is hard to machine highly dense MgB$_2$-based materials: they are brittle and can decompose (to form borane) in water or in humidity. But a recently developed (at the Institute for Superhard Materials – ISM NASU, Kiev, Ukraine) method of quick cutting even a highly dense material (without SC properties degradation) into items of practically any shape using waterless coolant opens new possibilities of the material applications.

The general view of blocks and rings synthesized from Mg and B (with additions of Ti and Ta) and the dependences of their critical current densities on magnetic field at various temperatures are shown in figure 26. The porosity not lowers than 98% and high critical currents are demonstrated by the materials sintered and synthesized at 2 GPa. The critical currents of MgB$_2$ synthesized by hot-pressing (30 MPa) technique are also rather high and large-sized blocks (porosity 22 - 8 % or even low) can be produced by this method. Using broaching for precompaction of raw materials followed by synthesis at atmospheric argon pressure or HIP (0.1 GPa), one can produce big long rings with critical currents shown in figure 26. Both synthesized at atmospheric pressure and HIP-produced for 2 h materials are rather porous (porosity 40 % and 38 %, respectively). The superconducting properties of the rings are somewhat lower than that of manufactured at high pressure and hot pressing, but using preliminarily broaching with a subsequent atmospheric or HIP synthesis large parts can be produced with superconducting and mechanical properties suitable for practical applications.

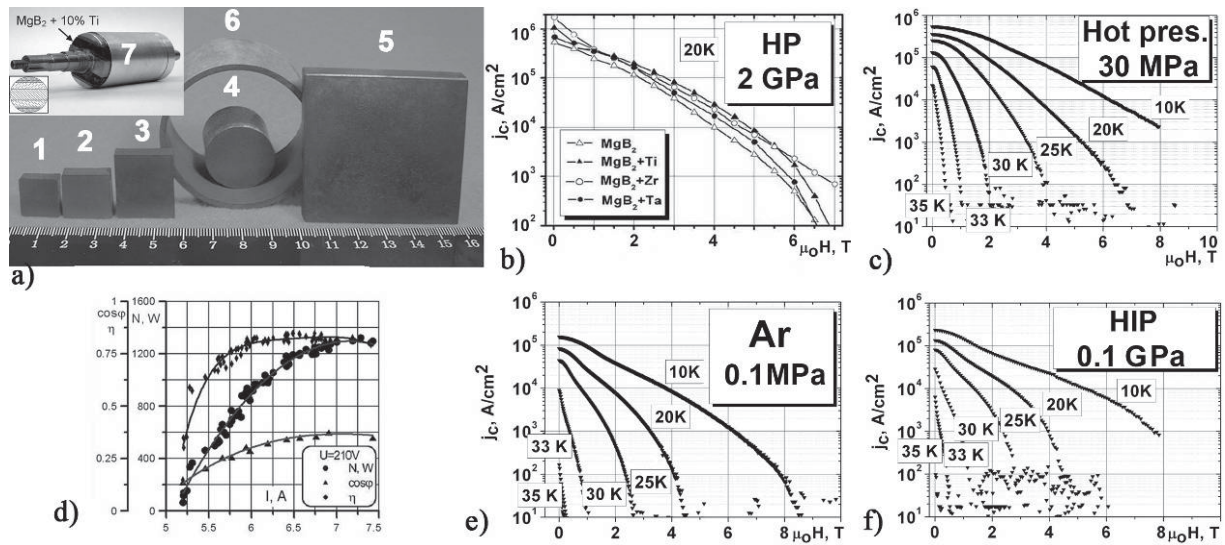

**Figure 26.** General view of blocks and rings synthesized from Mg and B taken in the stoichiometric mixture of MgB$_2$ with 10 % additions of Ti and Ta: 1–4 high-pressure synthesized at 800 °C, 2GPa, 1h (with additions of Ti); 5- hot pressed at 900 °C, 30 MPa, 1 h (with additions of Ta); 6 - broached and HIPed at 0.1 GPa, 900 °C, 1 h (with additions of Ti); 7 - rotor of reluctance motor with high-pressure (2 GPa) synthesized MgB$_2$ plates with 10% Ti addition (a); the generalized dependences of critical current density $j_c$ on magnetic field $\mu_0 H$ for high-pressure synthesized MgB$_2$ (at 2 GPa 750-900 °C for 1 h) without and with additions of Ta, Ti, Zr (2 and 10 wt %) at 20 K (b); dependences of $j_c$ on $\mu_0 H$ obtained by VSM for the items: hot-pressed block (c); rings synthesized under ambient argon pressure (e) and HIPed (f), respectively; characteristics of MgB$_2$-based SC rotor at 15-20 K (d) [75, 126].

The world first MgB$_2$ reluctance-type motor (a working temperature of 20 K) from high-pressure synthesized MgB$_2$ was produced in 2005 [75, 114, 126] (Figures 26a, d). The comparison study showed that the efficiency of a rotor manufactured from MgB$_2$ and tested at the same temperatures (15–20 K) as MT-YBCO did not differ essentially in performance. Highly dense MgB$_2$-based materials are rather hard and brittle and cannot be machined with water because of the decomposition to form poison boranes. Besides with the mentioned above method of quick cutting of dense MgB$_2$ (developed in ISM NANU) using waterless coolant to minimize the mechanical treatment, a special process aimed at manufacturing under high-pressure (2GPa) samples in the form of quadratic and rectangular blocks to be used in reluctance-type motors has been developed in ISM NANU as well.

High precision samples, for example needed for resonant measurements of elastic properties (group of three parallelepiped samples on left in Figure 27), can be prepared from HIPed material in the following way [104]. On the first step rotating diamond saw (60 grid, National Diamond) was used to cut plates from HIPed sample [104]. For a sample diameter of 30 mm it takes about 10 minutes to make a cut. Standard coolant (water plus rust resistance additive were used) at this stage. After this stage plates of magnesium diboride were attached to a metal magnetic plate and sliced into cubes of desirable sizes. Final operation to make sides parallel and edges sharp was performed using surface grinding machine with 60 grid diamond wheel for the rough removal of surface layer and 400 grid for finishing. Finally a thin deteriorated surface layer on magnesium diboride can be easily removed by dry polishing using fine sand paper producing mirror like surface (Figure 27).

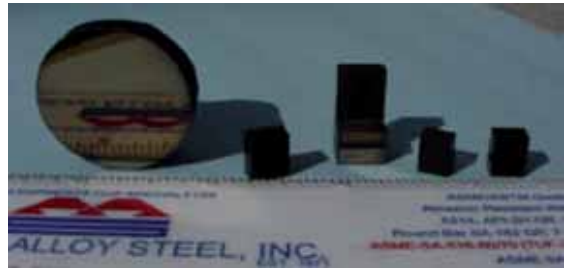

**Figure 27**. Examples of machined samples from HIPed magnesium diboride. Notice mirror like quality of polished sample on left and third from left. Three small parallelepipeds with sharp corners were machined to high accuracy RUS measurements of elastic constants [104].

Some prototypes of fault current limiters (FCLs) based on low-temperature superconductors operating at currents of 100–1500 A have been developed and successfully tested. However, the use of liquid helium as a cryogen makes maintenance of low-temperature superconducting devices rather complex and expensive. Therefore, the saving gained from the device application could be achieved only in networks of a very high power (of the order of 1 GVA). At the Budapest University of Technology and Economics the experimental study of the current limitation process has been conducted using devices of different configurations, which were the combination of high-temperature superconducting transformer equipped with the protector against the overloading, which combined the transformer and fault current limiter in the same device. A device of small size with copper primary and secondary windings and with one cylinder from MT-YBCO has been developed and constructed. In order to try for the similar purposes MgB$_2$-based bulk materials recently the MgB$_2$ cylinder of outer diameter = 21.3 mm, height = 14.1 mm, and wall thickness = 3.5 mm cut from high-pressure-synthesized samples has been tested in the model of inductive fault current limiter (Physics Department of Ben-Gurion University of the Negev, Israel). The tested MgB$_2$ cylinder showed high critical current density, the transition from superconductive state to normal and back was very fast, so this material can be successfully used for inductive current fault limiters. The cylinders of larger diameter can provide higher impedance change at the transition from the nominal regime to the limitation regime.

The world-first MgB$_2$-based motor (1.3 kW) from high pressure-high temperature synthesized MgB$_2$ at the ISM NASU, Kiev, Ukraine, has been developed in cooperation with MAI (Moscow, Russia) and IPHT (Jena, Germany). The SC rotor of the reluctance electromotor was manufactured from MgB$_2$ synthesized at 2 GPa, 800 $^o$C for 1 h with 10% of Ti addition (see insert on figure 26a). The comparative tests of the rotor with MT-YBCO at 15-20 K have shown that despite the lower SC transition temperature, the efficiency of the electromotor with MgB$_2$ superconductive elements proved to be of the same level as with the MT-YBCO elements.

## 5. Conclusions.

The recently developed manufacturing technologies of bulk MgB$_2$-based materials make it possible to produce polycrystalline nanostructural materials with near theoretical density and high superconducting and mechanical characteristics and items from these materials of sizes and shapes, which make them suitable for applications in cryogenic machines and devices like motors, fault current limiters, shielding screens, magnetic bearings, levitation transport, etc.

The highest for today critical current densities, $J_c$, at 20 K, at temperature which is considered to be the working temperature for MgB$_2$-based materials (boiling temperature of liquid hydrogen), were attained for bulk samples ($J_c$ estimated magnetically) with SiC, Ti and Zr addition high-pressure synthesized (at 2 GPa) and SiC-doped HIP- synthesized from mechanically alloyed Mg:B$_2$ =1:2 mixture (at 0.1 GPa) and malic acid (C$_4$H$_6$O$_5$) alloyed wires (transport $J_c$) synthesized (at 600$^o$C) after cold high-pressure densification (at 6.5 GPa) in contact with Nb. The highest $J$c at low temperatures (4.2-5 K) was demonstrated by bulk Ti Zr-dopped samples, tapes and wires nano-SiC and nano-C-doped (in magnetic fields 6-16 T) prepared in contact with Ti or Nb.

The mechanism of Ti, Ta, Zr influence on an increase of critical current density with a high probability is connected with: (1) possibility of these elements to absorb hydrogen at low synthesis temperatures, thus (i) promoting the formation of higher borides (mainly with near MgB$_{12}$ stoichiometry in the case of high-pressure (2 GPa) synthesized materials), dispersed inclusions of which can be responsible for pinning; (ii) improving connectivity by decreasing the porosity, reducing amount of cracks and preventing the formation of MgH$_2$; (2) under high synthesis temperatures their presence may enhance an oxygen segregation to form oxygen-enriched Mg-B-O inclusions, which can also be responsible for pinning and thus cleaning grain boundaries of MgB$_2$ from oxygen and improving grain-boundary pinning. While there exists an opinion that Ti, Ta, Zr can form "a network" from TiB$_2$ or ZrB$_2$ grains, the size of which is about one unit cell and that such a "network" refines the structure of MgB$_2$ and thus, leads to an increase in $J_c$. It should be noted that the investigation of highly dense high-pressure synthesized materials, which demonstrated very high $J_c$ values, does not support the above point of view.

In the case of SiC and C doping, it has been revealed that the SiC doping allows the C substitution and MgB$_2$ formation to take place simultaneously at low temperatures. The C substitution enhances $H_{c2}$, while the defects, small grain size and nanoinclusions induced by the C incorporation and low temperature processing are responsible for the improvement in $Jc$. A unified mechanism has been proposed, according to which the optimal doping effect can be achieved when the C substitution and MgB$_2$ formation take place at the same time at low temperatures: C substitution for B induces disorder in lattice sites, increases the resistivity and, hence, an enhancement in $H_{c2}$, while the C substitution together with low-temperature processing results in the reduction of grain sizes, fluctuations in $T_c$, extra defects, and embedded inclusions that enhance the flux pinning. Thus, SiC doping takes advantage of both C substitution and low-temperature processing. But the amount of C embedded into the MgB$_2$ structure should not be too large so that the $T_c$ should not be decreased essentially. It has been shown that for MgB$_2$–$n$D$_x$ ($n$D – nanodiamonds) superconducting transition temperature ($T$c) is not affected by $x$ up to $x$ = 0.05. In addition to this proposed unified mechanism: the probability of SiC influence on oxygen segregation in the MgB$_2$ structure at high synthesis temperature similar to that observed in the case of Ti or Ta addition should not be excluded. Probably the high critical current densities of wires alloyed by SiC and C-containing compounds can be explained not only due to the specially induced alloyed additions to MgB$_2$ but as well due to the contact with hydrogen-absorbing elements, such as, for example, Ti and Nb.

The fact that rather high critical current densities (commensurable with that for the materials with additions) can be attained in materials produced only from Mg and B or from MgB$_2$ is support the idea concerning the primary importance for high superconducting characteristic of MgB$_2$-based materials of the character or manner of oxygen (oxygen enriched Mg-B-O inclusions, in particular), boron (or higher borides), hydrogen, and carbon admixture (which probably present in the starting materials) distribution in the material. Some improvement of properties of undoped materials can be attained by the mechanical activation of the initial powders, adding some excessive Mg to the initial mixture, synthesis in magnetic field or under elevated or high pressure, by presintered compacting of the initial mixture under high pressure, etc.

Usually different synthesis or sintering temperatures should be used for attaining high critical current densities in high and low magnetic fields as well as at different temperatures (at least in the 10-35 K temperature range).

The possibility to attain a comparatively high critical current density in a material with near MgB$_{12}$ matrix points to the probability that the MgB$_{12}$ phase has the SC properties (with $T_c$ about 37 K) or that for a flow of comparatively high percolation current it is surprisingly small amount of MgB$_2$ phase in the material enough to be present. The existed uncertainty in connection with the determination of MgB$_{12}$ structure (orthorhombic or hexagonal) and a mosaic structure of grains of phase with near MgB$_{12}$ stoichiometry, which are formed in polycrystalline materials, still leave the question open, despite the mentioned in literature fact that for the orthorhombic MgB$_{12}$ single crystal preliminarily measurements showed the absence of SC down to 2 K.

## Acknowledgements

The author is very grateful to Prof. Wolfgang Gawalek, Dr. Wilfried Goldacker, Prof. Rene Flükiger, and Dr. Michael Eisterer for the support of this work and fruitful discussions.

## References

[1] Nagamatsu J, Nakagawa N, Muranaka T, Zenitani Y and Akimitsu J 2001 *Nature* **410** 63
[2] Bud'ko S L, Lapertot G, Petrovic C, Cunningham C E, Anderson N and Canfield P C 2001 *Phys. Rev. Lett.* **86** 1877
[3] Buzea C and Yamashita T 2001 *Supercond. Sci. Technol.* **14** 115
[4] Eisterer M 2007 *Supercond. Sci. Technol.* **20** 47
[5] Bouquet F, Fisher R A, Phillips N E, Hinks D G and Jorgensen J D 2001 *Phys. Rev. Lett.* **87** 047001
[6] Bouquet F, Wang Y, Sheikin I, Plackowski T, Junod A, Lee S and Tajima S 2002 *Phys. Rev. Lett.* **89** 257001
[7] Bugoslavsky Y et al 2002 *Supercond. Sci. Technol.* **15** 526
[8] Giubileo F, Roditchev D, Sacks W, Lamy R, Thanh D X, Klein J, Miraglia S, Fruchart D, Marcus J and Monod Ph 2001 *Phys. Rev. Lett.* **87** 177008
[9] Gonnelli R S, Daghero D, Ummarino G A, Stepanov V A, Jun J, Kazakov S M and Karpinski J 2002 *Phys. Rev. Lett.* **89** 247004
[10] Iavarone M et al 2002 *Phys. Rev. Lett.* **89** 187002
[11] Schmidt H, Zasadzinski J F, Gray K E and Hinks D G 2002 *Phys. Rev. Lett.* **88** 127002
[12] Szab.o P, Samuely P, Kaˇcmarˇc.ık J, Klein T, Marcus J, Fruchart D, Miraglia S, Marcenat C and Jansen A G M 2001 *Phys. Rev. Lett.* **87** 137005
[13] Manzano F, Carrington A, Hussey N E, Lee S, Yamamoto A and Tajima S 2002 *Phys. Rev. Lett.* **88** 047002
[14] Chen X K, Konstantinovi.c M J, Irwin J C, Lawrie D D and Franck J P 2001 *Phys. Rev. Lett.* **87** 157002

[15] Tsuda S, Yokoya T, Kiss T, Takano Y, Togano K, Kito H, Ihara H and Shin S 2001 *Phys. Rev. Lett.* **87** 177006
[16] Tu J J, Carr G L, Perebeinos V, Homes C C, Strongin M, Allen P B, Kang W N, Choi E-M, Kim H-J and Lee S-I 2001 *Phys. Rev. Lett.* **87** 277001
[17] Moshchalkov V V, 2009 *Abstracts of Six International Conference in School Format on Vortex Matter in Nanostructured Superconductors* 54
[18] Daghero D, Gonnelli R S, Ummarino G A, Stepanov V A, Jun J, Kazakov S M and Karpinski J 2003 *Physica C* **385** 255
[19] Gonnelli R S, Daghero D, Ummarino G A, Stepanov V A, Jun J, Kazakov S M and Karpinski J 2003 *Supercond. Sci. Technol.* **16** 171
[20] Iavarone M et al 2003 *Supercond. Sci. Technol.* **16** 156
[21] Collings E W, Sumption M D, Bhatia M, Susner M A and Bohnenstiehl S D 2008 *Supercond. Sci. Technol.* **21** 103001
[22] Rowell J M 2003 *Supercond. Sci. Technol.* **16** 17
[23] Rowell J M et al 2003 *Appl. Phys. Lett.* **83** 102
[24] Gurevich A et al 2004 *Supercond. Sci. Technol.* **17** 278
[25] Shields T C, Kawano K, Holdom D and Abell J S 2002 *Supercond. Sci. Technol.* **15** 202
[26] Finnemore D K, Ostenson J E, Bud'ko S L, Lapertot G and Canfield P C 2001 *Phys. Rev. Lett.* **86** 2420
[27] Polyanskii A A, Gurevich A, Jiang J, Larbalestier D C, Bud'ko S L, Finnemore D K, Lapertot G and Canfield P C 2001 *Supercond. Sci. Technol.* **14** 811
[28] Kambara M, Hari Babu N, Sadki E S, Cooper J R, Minami H, Cardwell D A, Campbell A M and Inoue I H 2001 *Supercond. Sci. Technol.* **14** L5
[29] Kawano K, Abell J S, Kambara M, Hari Babu N and Cardwell D A 2001 *Appl. Phys. Lett.* **79** 2216
[30] Larbalestier D C et al 2001 *Nature* **410** 186
[31] Cambel V, Fedor J, Gregušova D, Kovač P and Hušek I 2005 *Supercond. Sci. Technol.* **18** 417
[32] Drozd A M, Gabovich V A, Gierłowski P, Pekała M and Szymczak H 2004 *Physica C* **402** 325
[33] Eisterer M, Zehetmayer M, Tönies S, Weber H W, Kambara M, Hari Babu N, Cardwell D A and Greenwood L R 2002 *Supercond. Sci. Technol.* **15** L9
[34] Dew-Hughes D 1974 *Phil. Mag.* **30** 293
[35] Campbell A M and Evetts J E 1972 *Critical Currents in Superconductors*
[36] Collings E W 1986 *Applied Superconductivity Metallurgy and Physics of Titanium Alloys* **2** Part 1
[37] Birajdar B, Peranio N and Eibl O 2008 *Supercond. Sci. Technol.* **21** 073001
[38] Prikhna T A, Gawalek W, Savchuk Ya M, Surzhenko A B, Zeisberger M, Moshchil V E, Dub S N, Melnikov V S, Sergienko N V, Habisreuther T, Litzkendorf D, Abell S, Nagorny P A 2003 *IEEE transactions on Applied Superconductivity* **13** 3506
[39] Prikhna T A, Gawalek W, Savchuk Ya M, Sergienko N V, Moshchil V E, Wendt M, Zeisberger M, Habisreuther T, Sverdun V B, Dou S X, Dub S N, Melnikov VS, Schmidt Ch, Dellith J and Nagorny P A 2006 *Journal of Physics: Conference Series* **43** 496
[40] Prikhna T A, Gawalek W, Savchuk Ya M, Habisreuther T, Wendt M, Sergienko N V, Moshchil V E, Nagorny P, Schmidt Ch, Dellith J, Dittrich U, Litzkendorf D, Melnikov V S and Sverdun V B 2007 *Supercond. Sci. Technol.* **20** 257
[41] Prikhna T A, Gawalek W, Tkach V M, Danilenko N I, Savchuk Ya M, Dub S N, Moshchil V E, Kozyrev A V, Sergienko N V, Wendt M, Melnokov V S, Dellith J, Weber H, Eisterer M, Schmidt Ch, Habisreuther T, Litzkendorf D, Vajda J, Shapovalov A P, Sokolovsky V, Nagorny P A, Sverdun V B, Kosa J, Karau F, Starostina A V (*in press Journal of Physics*).

[42] Prikhna T, Gawalek W, Novikov N, Savchuk Ya, Moshchil V, Sergienko N, Wendt M, Dub S, Melnikov V, Surzhenko A, Litzkendorf D, Nagorny P, Schmidt Ch 2003 *Proceedings of CIMTEC* 2002*, and cond-mat/* 0204362
[43] Prikhna T A, Gawalek W, Savchuk Ya M, Moshchil V E, Sergienko N V, Habisreuther T, Wendt M, Hergt R, Schmidt Ch, Dellith J, Melnikov V S, Assmann A, Litzkendorf D, Nagorny P A 2004 *Physica C* **402** 223
[44] Schmitt R 2006 *Dissertation am Institut fur Anorganische Chemie, University of Tübingen*
[45] Schmitt R, Glaser J, Wenzel T, Nickel K G, Meyer H-J 2006 *Physica C* **436** 38
[46] Wenzel T, Nickel K.G, Glaser J, Meyer H-J, Eyidi D, Eibl O 2003 *Physica Status Solidi* **198** 374.
[47] Birajdar B, Braccini V, Tumino A, WenzeT, Eibl O and GrassoG 2006 *Supercond. Sci. Technol.* **19** 916
[48] Giunchi G, Ripamonti G, Raineri S, Botta D, Gerbaldo R, Quarantiello R 2004 *Supercond. Sci. Technol.* **17** S583
[49] Markovski L Y,. Kondrashev Y D, Kapuatovskaya G V, Gen J 1955 *Chem. USSR* **25** 409
[50] Turkevich V, Kulik O, Icenko P, Sokolov A, Lucenko A and Vashenko A 2003 *Superhard materials* **1** 9
[51] Turkevich V Z, Prikhna T A and Kozyrev A V 2009 *High Pressure Research* **29:1** 87
[52] Brutti S, Colapietro M, Calducci G, Barba L, Manfrinetti P, Palenzona A 2002 *Intermetallics* **10** 811.
[53] Volker Adasch Synthese 2005 *Dissertation, zur Erlangung des akademischen Grades*
[54] Adasch V, Hess K-U, Ludwig T, Vojteer N, Hillebrecht H 2006 *Journal of Solid State Chemistry* **179** 2916
[55] Savchuk Y M, Prikhna T O, Moshchil V E, Sergienko N V, Nagorny P A, Melnikov V S, Dub S M, Gawalek W, Surzhenko O B, Wendt M 2003 *Journal Of Superhard Materials* **25** 44
[56] Prikhna T, Gawalek W, Savchuk Y, Sergienko N, Wendt M, Habisreuther T, Moshchil V, Mamalis A, Noudem J, Chaud X, Turkevich V, Nagorny P, Kozyrev A, Dellith J, Shmidt C, Litzkendorf D, Dittrich U and Dub S 2008 *Journal of Optoelectronics and Advanced Materials* **10** 1017
[57] Dou S X et al 2007 *Phys. Rev. Lett*. **98** 097002
[58] Ferrando V et al 2007 *J. Appl. Phys.* **101** 043903
[59] Wang S-F, Liu Z, Zhou Y-L, Zhu Y-B, Chen Z-H, Lu H-B, Cheng B-L and Yang G-Z 2004 *Supercond. Sci. Technol.* **17** 1126
[60] Jones M E, Marsh R E, Am J 1953 *Chem. Soc.* **76** 1434.
[61] Yu R C, Li S C, Wang Y Q, Kong X, Zhu J L, Li F Y, Liu Z X, Duan X F, Zhang Z and Jin C Q 2001 *Physica C* **363** 84
[62] Klie R F, Zhu Y, Schneider G and Tafto J 2003 *Appl. Phys. Lett.* **82** 4316
[63] Keast V J 2001 *Appl. Phys. Lett.* **79** 3491
[64] Eom C B, Lee M K, Choi J H, Belenky L J, Song X, Cooley L D, Naus M T, Patnaik S, Jiang J, Rikel M, Polyanskii A, Gurevich A, Cai X Y, Bu S D, Babcock S E, Hellstrom E E, Larbalestier D C, Rogado N, Regan K A, Hayward M A, He T, Slusky J S, Inumaru K, Haas M K and Cava R J 2001 *Nature* **411** 558.
[65] Klie R F, Idrobo J C, Browning N D, Serquis A, Zhu Y T, Liao X Z and Mueller F M 2002 *Appl. Phys. Let* **80** 213970
[66] Prikhna T A, Gawalek W, Savchuk Ya M, Kozyrev A V, Wendt M, Melnikov V S, Turkevich V Z, Sergienko N V, Moshchil V E, Dellith J, Ch. Shmidt, Dub S N, Habisreuther T, Litzkendorf D, Nagorny P A, Sverdun V B, Weber H W, Eisterer M, Noudem J and Dittrich U 2009 *IEEE Transactions on Applied Superconductivity* **19** 2780
[67] Liao X Z, Serquis A C, Zhu Y T, Huang J Y, Civale L, Peterson D E, Mueller F M and Xu H 2003 *Journal of Applied Physics* **93** 6208

[68] Liao X Z, Serquis A, Zhu Y T, Huang J Y, Civale L, Peterson D E, Mueller F M, Xu H *F cond-mat/* 0212571
[69] Prikhna T, Gawalek W, Savchuk Ya, Kozyrev A, Wendt M, Dellith J, Goldacker W, Dub S, Sergienko N, Habisreuther T, Moshchil V, Dittrich U, Karau W, Noudem J, Schmidt Ch, Turkevich V, Litzkendorf D, Weber H, Eisterer M, Melnikov V, Nagorny P, Sverdun V *Journal of the KIASC (Korea Institute of Applied Superconductivity and Cryogenics). ICEC22-ICMC2008 (in press)*
[70] Prikhna T, Gawalek W, Mamalis At, Savchuk Ya, Sergienko N, Wendt M, Habisreuther T, Noudem J, Chaud X, Moshchil V, Turkevich V, Nagorny P, Kozyrev A, Dellith J, Shmidt Ch, Litzkendorf D, Dittrich U, Dub S *2007 Fifth Japanese-Mediterranean Workshop on Applied Electromagnetic Engineering for Magnetic Superconducting and Nano Materials* 7
[71] Prikhna T A, Gawalek W, Savchuk Ya M, Moshchil V E, Sergienko N V, Surzhenko A B, Wendt M, Dub S N, Melnikov V S, Schmidt Ch, Nagorny P A 2003 *Physica C* **386** 565
[72] Eyidi D, Eibl O, Wenzel T, Nickel K G, Giovannini M and Saccone A 2003 *Micron* **34** 85
[73] Birajdar B, Braccini V, Tumino A, Wenzel T, Eibl O and Grasso G 2006 *Supercond. Sci. Technol.* **19** 916
[74] Haessler W, Birajdar B, Gruner W, Herrmann M, Perner O, Rodig C, Schubert M, Holzapfel B, Eibl O and Schultz L 2006 *Supercond. Sci. Technol.* **19** 512
[75] Prikhna T A, Gawalek W, Savchuk Ya, Sergienko N V, Moshchil V E, Wendt M, Habisreuther T, Dub S N, Melnikov V S, Kozyrev A V, Schmidt Ch, Dellith J, Litzkendorf D, Nagorny P A, Dittrich U, Sverdun V B, Kovalev L K, Penkin V T, Goldacker W, Rozenberg O A Noudem J 2008 *Journal of Physics* **97** 012022
[76] Gunji S, Kanimura H 1996 *Phys. Rev. B* **54** 13665
[77] Soga K, Oguri A, Araake S, Kimura K, Terauchi M, Fujiwara A 2004 *J. Solid State Chem.* **177** 498
[78] Haigh S, Kovac P, Prikhna T A, Savchuk Ya M, Kilburn M, Salter C, Hutchison J and Grovenor C 2005, *Supercond. Sci. Technol.* **18** 1190
[79] Grovenor C R M, Goodsir L, Salter C J, Kovac P and Husek I 2004 *Supercond. Sci. Technol.* **17** 479
[80] Dou S. X, Soltanian S., Horvat J., et al., 2002 *Appl. Phys. Lett.* **81** 3419
[81] Zhao Y., Feng Y.,. Cheng C.H et al., 2001 *Appl. Phys. Lett.* **79** 1154
[82] Goto D., Machi T., Zhao Y. et al., 2003 *Physica C*, **392–396**, 272.
[83] Suo H L, Beneduce C, Dhalle M, Musolino N, Genoud J Y and Flukiger R 2001, cond-mat/0106341.
[84] Kim K H P, Kang W N, Kim M S, Jung C U, Kim H J, Choi E M, Park M S and Lee S I 2001 cond-mat/0103176
[85] Dhalle M, Toulemonde P, Beneduce C, Musolino N, Decroux M and Flukiger R 2001 *cond-mat/* 0104395
[86] Flükiger R., Hossain M. S. A., Senatore C., 2009 *cond-mat /*0901.4546 (for SUST)
[87] Maeda M., Zhao Y., Dou S. X, Nakayama Y., Kawakami T., Kobayashi H., Kubota Y. 2008 *Supercond. Sci. Technol.* **21** 032004
[88] Jiang C H, Dou S X, Cheng Z X and Wang X L 2008 *Supercond. Sci. Technol.* **21** 065017
[89] Mikheenko P, Martınez E, Bevan A, Abell J S, MacManus-Driscoll J L, 2007 *Supercond. Sci. Technol.* **20** 264
[90] Vajpayee A., Huhtinen H, Awana V P S, Gupta1 A., Rawat R., Lalla N P, Kishan H, Laiho R, Felner I, Narlikar A V 2007 *Supercond. Sci. Technol.* **20** 155
[91] Zeng R, Lu L, Wang J L, Horvat J, Li W X, Shi D Q, Dou S X, Tomsic M, Rindfleisch M 2007 *Supercond. Sci. Technol.* **20** L43
[92] Hur J M, Togano K, Matsumoto A, Kumakura H, Wada H and Kimura K 2008 *Supercond. Sci. Technol.* **21** 032001.

[93]   Ma Y., Xu A., Li X., Zhang X., Awajl S., Watanabe K., 2006 *Japanese Journal of Applied Physics* **45** L493
[94]   Matsumoto A, Kitaguchi H and Kumakura H 2008 *Supercond. Sci. Technol.* **21** 065007
[95]   Zhao Y., Feng Y., Huang D.X., Machi T., Cheng C.H., Nakao K., Chikumoto N., Fudamoto Y., Koshizuka N., Murakami M. 2002  *Physica C* **378–381** 122
[96]   Dou S.X., Shcherbakova O., Yeoh W.K., Kim J.H., Soltanian1 S., Wang X.L., Senatore C., Flukiger R., Dhalle M., Husnjak O., and Babic E. 2007 cond-mat/0701391
[97]   Horhager N, Eisterer M, Weber H W, Prikhna T, Tajima T, Nesterenko V F, 2006 *Institute of Physics Publishing Journal of Physics: Conference Series, 7th European Conference on Applied Superconductivity* **43** 500
[98]   Ma Y, Zhang X, Nishijima G., Watanabe K., Awaji S., Bai X. 2006 *Appl. Phys. Lett*. **88** 072502
[99]   Soltanian S., Horvat J, Wang X L, Munroe P, Dou S X 2003    *Physica C* **390** 185
[100]  Susner M A, Sumption M D, Bhatia M, Peng X, Tomsic M J, Rindfleisch M A, Collings E W 2007 *Physica C* **456** 180
[101]  Hossain M S A, Senatore C, Fluekiger R, M A Rindfleisch, M J Tomsic, J H Kim and S X Dou 2009*, Supercond. Sci. Technol.* **22** 095004
[102]  Hässler W, Herrmann M, Rodig C, Schubert M, Nenkov K and Holzapfel B 2008 *Supercond. Sci. Technol.* **21** 062001
[103]  Braccini V et al 2007 *IEEE Trans. Appl. Supercond.* **17** 2766
[104]  Nesterenko Vitali F,  2002 *cond-mat* /0212543
[105]  Prikhna T, Gawalek W, Savchuk Ya, Tkach V, Danilenko N, Wendt M, Dellith J, Weber H, Eisterer M, Moshchil V, Sergienko N, Kozyrev A, Nagorny P, Shapovalov A, Melnikov V, Dub S, Litzkendorf D, Habisreuther T, Schmidt Ch, Mamalis A, Sokolovsky V, Sverdun V, Karau F, Starostina A, 2010 *(in press Physica C)*
[106]  Kováč P, Hušek I, Melišek T and Holúbek T 2007 *Supercond. Sci. Technol.* **20** 771.
[107]  Prikhna T A, Eisterer M, Savchuk Ya, Tomsic M et al. (in press).
[108]  Schlachter S et al 2007 *Paper M1H02 presented at ICMC*
[109]  Wang X L et al., 2004 *Physica C* **408-410** 63
[110]  Eisterer M, Robert Schoppl K, Weber H W, Sumption M D and Bhatia M 2007 *IEEE Trans. Appl. Supercond.* **17** 2814
[111]  Prikhna T, Novikov N, Savchuk Ya, Gawalek W, Sergienko N,  Moshchil V, Wendt M, Melnikov V, Dub S, Habisreuther T,  Schmidt Ch, Dellith J, Nagorny P,  2005 *Innovative Superhard Materials and Sustainable Coatings for Advanced Manufacturing. Edited by Jay Lee, Nikolay Novikov.  NATO Science Series. II. Mathematics, Physics and Chemistry* **200** 81
[112]  Lee S, Masui T, Yamamoto A, Uchiyama H, Tajima S  2003 *Physica C* **397** 7
[113]  Deemyad S, Tomita T, Hamlin J J, Beckett B R, Schilling J S, Hinks D G, Jorgensen J D, Lee S, Tajima S 2003 *Physica C* **385** 105
[114]  Prikhna T, Gawalek W, Savchuk Ya, Sergienko N, Moshchil V, Dub S, Sverdun V, Kovalev L, Penkin V, Zeisberger M, Wendt M, Fuchs G, Habisreuther T, Litzkendorf D, Nagorny P, Melnikov V  2007 *Physica C* **460–462**  595
[115]   Evans A G, Charles E A 1976 *J.Am. Ceram. Soc.* **59**, № 7-8 371
[116]  A. P. Malozemoff, Application of high temperature superconductors, in Ceramic materials in Energy Systems for Sustainable Development, Word Academy of Ceramics, Ed. L.J. Gauckler, Techna group –monographs in materials and society, vol. 8, 279- 295, 2009.
[117]  Murakami M. 2007  *Int. J. Appl. Ceram. Technol.,* **4 [3]** 225
[118]  Vajda I and Farkas L 2004  *Advanced Studies on Superconducting Engineering", second edition Ed:, Published by the Supertech Laboratory Budapest University of Technology and Economics* 308
[119]  Shim S H, Shim K B, Yoon J W 2005 *J. Am. Ceram. Soc.* **88**  858

[120] Song K J, Park C, Kim S W, Ko R K, Ha H S, Kim H S, Oh S S, Kwon Y K, Moon S H, Yoo S I 2005 *Physica C* **426** 588
[121] Schmidt J, Schnelle W, Grin Y, Kniep R 2003 *Solid State Sci.* **5** 535
[122] Cao G, Locci A M, Orru` R, 2005 *Patent PCT/EP2005/052857*
[123] Locci A M, Orru` R, Cao G, Sanna S, Congiu F, Concas G 2006 *AIChE J.* **52** 2618
[124] Aldica G, Badica P, Groza J R, 2007 *J. Optoelectron. Adv. Mater.* **9** 1742
[125] Orru` R, Licheri R, Locci A M, Cincotti A, Cao G 2009 *Materials Science and Engineering R* **63** 127
[126] Kovalev L K, Ilushin K V, Penkin V T, Kovalev K L, Poltavets V N, Koneyev S MA, ModestovKA, Gawalek W, Prikhna T A, Akimov I I 2006 *Journal of Physics: Conference Series* **43** 792